\documentclass[12pt, preprint]{emulateapj}

\usepackage{lscape} 

\shortauthors{Shi et al.}

\begin{document}

\title{Aromatic Features in AGN: Star-Forming Infrared Luminosity Function of AGN Host Galaxies}

\author{Yong Shi\altaffilmark{1}, Patrick Ogle\altaffilmark{2}, George H. Rieke\altaffilmark{1} 
, Robert Antonucci\altaffilmark{3}, Dean C. Hines\altaffilmark{4}, 
Paul S. Smith\altaffilmark{1}, Frank J. Low\altaffilmark{1}, Jeroen Bouwman\altaffilmark{5}, 
Christopher Willmer\altaffilmark{1}}

\altaffiltext{1}{Steward Observatory, University of Arizona, 933 N Cherry Ave, Tucson, AZ 85721, USA}
\altaffiltext{2}{Spitzer Science Center, California Institute of Technology, Mail Code 220-6, Pasadena, CA 91125}
\altaffiltext{3}{Physics Department, University of California, Santa Barbara, CA 93106}
\altaffiltext{4}{Space Science Institue 4750 Walnut Street, Suite 205, Boulder, Colorado 80301}
\altaffiltext{5}{Max-Planck-Institut fu$\:$r Astronomie, D-69117 Heidelberg, Germany}

\begin{abstract}

We describe observations  of aromatic features at 7.7  and 11.3 $\mu$m
in  AGN of  three  types including  PG,  2MASS and  3CR objects.   The
feature  has been  demonstrated to  originate predominantly  from star
formation. Based  on the aromatic-derived star  forming luminosity, we
find that the  far-IR emission of AGN can be  dominated by either star
formation  or nuclear  emission;  the average  contribution from  star
formation  is around  25\% at  70  and 160  $\mu$m.  The  star-forming
infrared luminosity  functions of the  three types of AGN  are flatter
than  that  of field  galaxies,  implying  nuclear  activity and  star
formation tend  to be enhanced together.   The star-forming luminosity
function is also  a function of the strength  of nuclear activity from
normal  galaxies  to the  bright  quasars,  with luminosity  functions
becoming flatter  for more intense nuclear  activity.  Different types
of AGN  show different  distributions in the  level of  star formation
activity, with 2MASS $>$ PG $>$ 3CR star formation rates.

\end{abstract}                                                    

\section{INTRODUCTION} 

The  interplay  between  supermassive  black holes  (SMBHs)  and  star
formation  is  now  recognized  as  a critical  ingredient  in  galaxy
evolution, as  demonstrated by the correlations  between the blackhole
masses and  the bulge properties of their  host galaxies ($M$-$\sigma$
relation)  \citep{Kormendy95,  Magorrian98, Gebhardt00,  Ferrarese00}.
However, because the star formation rate (SFR) is difficult to measure
around active  galactic nuclei  (AGN), we are  unable to  answer basic
questions about the interrelations  between the two processes: in what
star-forming  environments does  AGN  activity tend  to be  triggered?
Does feedback from one process trigger or quench another?

Models that  involve the galaxy merging process  and AGN feedback
simulate        the       $M$-$\sigma$        relation       successfully
\citep[e.g.][]{DiMatteo05}.  The  theoretical picture of  the ``cosmic
cycle'' of galaxy  evolution \citep[e.g.][]{Hopkins06} connects galaxy
mergers, starbursts and nuclear  accretion.  Galaxy mergers induce gas
inflow  producing starbursts  and  obscured quasar  activity.  As  the
quasar feedback starts to heat and expel the circumnuclear medium, the
nuclear  activity   becomes  visible  as   optically  bright  quasars.
Eventually, the  quasar activity and starbursts are  terminated as the
gas and dust are more thoroughly expelled.  In this scenario, the time
histories  of the  star formation  and nuclear  accretion  through the
merging  process are  two fundamental  physical  properties underlying
many  observations  \citep[e.g.][]{Granato04, Springel05,  Hopkins06}.
However, current  observations only provide  detailed understanding of
star formation in normal galaxies, not in those dominated by luminous AGN.

While the  near- and mid-IR  emission of AGN  arise from hot  and warm
dust  heated  by  nuclear emission  \citep[e.g.][]{Polletta00,  Shi05,
Hines06, Jiang06}, the heating  mechanism of the cold dust responsible
for the far-IR emission still remains ambiguous \citep[See][]{Haas03}.
As  suggested  by  numerical  simulations  \citep{Chakrabarti06},  the
contribution  of  the AGN  to  the  far-IR  emission may  characterize
different  evolutionary  stages.  Insights  into  the far-IR  emission
mechanism  can  also  constrain  the structure  of  the  circumnuclear
material and its  evolution with redshift \citep[See][]{Ballantyne06}.
It  is also  critical  to understand  the  energy budget  of many  AGN
revealed in deep IR surveys \citep[e.g.][]{Alonso-Herrero06, Donley05,
Donley07}, which are faint in the optical, or even in X-ray bands, and
whose main energy output resides at infrared wavelengths.  Progress on
these topics requires the ability to identify the contribution of star
formation to the IR emission.

Although  the  commonly used  star-formation  tracers  (the total  UV,
H$\alpha$ and IR emission) may be contaminated severely by the nuclear
emission, there are  several alternatives to estimate the  SFR in AGN,
such as the extended UV emission, extended mid-IR emission, and narrow
metal emission lines.   The extended UV emission can  be observed with
high-resolution  telescopes such as  {\it HST}.   However, due  to the
large brightness  contrast between type 1  AGN and the  host galaxy in
the UV, this  method is limited to  type 2 AGN, and even  for them the
scattered nuclear  UV emission may  be significant \citep{Zakamska06}.
Extended mid-IR emission has been  used to estimate the SFR for nearby
Seyfert  galaxies  \citep[e.g.][]{Maiolino95}.   Due  to  the  limited
angular  resolution of  infrared telescopes,  it becomes  difficult to
resolve the AGN  from the circumnuclear star formation  for objects at
$z>$0.05  (0.5$''$=500pc).   Estimating  the  SFR  with  narrow  metal
emission lines is  difficult because they are contaminated  by the AGN
narrow emission  line region.  In  addition, this method  suffers from
other problems, for example, the [OII]$\lambda$3727 flux of PG quasars
indicates a very low SFR  \citep{Ho05}, which is inconsistent with the
abundant  molecular gas  in these  objects  and possibly  a result  of
under-estimating  the  amount  of  extinction  of  the  emission  line
\citep{Schweitzer06}.

In  this  study,  we  employ  the mid-infrared  aromatic  features  to
quantify the SFR  in AGN host galaxies.  These  features are prominent
at 3.3, 6.2,  7.7, 8.6, 11.3 and 12.7  $\mu$m \citep{Gillett73}.  They
are  seen  in various  Galactic  environments  including HII  regions,
diffuse interstellar clouds, planetary nebulae, reflection nebulae and
photodisassociation  regions  (PDRs)   and  in  extragalactic  objects
\citep[for  a  review, See][]{Tielens99}.   The  aromatic emission  in
normal  star-forming   galaxies  is   similar  to  that   in  Galactic
star-forming regions  \citep[e.g.][]{Genzel98, Clavel00}, with  a well
understood correlation  to the SFR  \citep[e.g.][]{Roussel01, Dale02}.
The aromatic  features in active  galaxies have much  lower equivalent
width  (EW)  than  in  star-forming  galaxies  \citep[e.g.][]{Roche91,
Clavel00}, implying  the destruction of  the aromatic carriers  by the
harsh nuclear radiation  or the inability of the  nuclear radiation to
excite the aromatic features.  Evidence for excitation of the aromatic
features  by star formation  in active  galaxies comes  from spatially
resolved  mid-IR  spectra  of  nearby  examples,  where  the  observed
aromatic  emission  is mainly  from  the disk  \citep[e.g.][]{Cutri84,
Desert88, Voit92, Laurent00, LeFloch01}.  Various infrared diagnostics
have  been  developed  based  on  a correlation  of  aromatic  feature
strength with star forming  activity to discriminate the power sources
(star  formation  versus   nuclear  activity)  for  luminous  infrared
galaxies         (LIRGs;         $L_{IR}>10^{11}$         L$_{\odot}$)
\citep[e.g.][]{Genzel98,   Laurent00,   Tran01,  Peeters04}.    Direct
measurements of the aromatic features in a small PG quasar sample have
been  carried out by  \citet{Schweitzer06}  to study  the  quasar  far-IR
emission mechanism.

In   this   paper,   we   present  $Spitzer$   Infrared   Spectrograph
\citep[IRS;][]{Houck04} low-resolution  spectra for a  large sample of
AGN.  \S~2 describes the sample, the data reduction, the extraction of
the  features at  7.7 and  11.3 $\mu$m  and the  determination  of the
associated  uncertainties.   In  \S~3,  we provide  evidence  for  the
star-formation excitation  of the  aromatic feature in  these objects.
In \S~4, we  estimate the conversion factor from  the aromatic flux to
the total IR  flux. \S~5 discusses the origin  of AGN far-IR emission.
In \S~6, we  construct the luminosity function of the  SFR in AGN host
galaxies and discuss its  implication for AGN activity.  \S~7 presents
our  conclusions.   Throughout this  paper,  we  assume $H_{0}$=70  km
s$^{-1}$ Mpc$^{-1}$, $\Omega_{0}$=0.3 and $\Omega_{\Lambda}$=0.7.

\section{DATA AND ANALYSIS}
\subsection{Sample}

The sample  in this  paper is composed  of objects derived  from three
parent   samples   selected   by   different   techniques:  
optically-selected  Palomar-Green (PG) quasars  \citep{Schmidt83}; the
Two-Micron  All Sky  Survey (2MASS)  quasars \citep{Cutri01};  and 3CR
radio galaxies and quasars \citep{Spinrad85}.  PG quasars are selected
at $B$ band to have blue $U$-$B$ color, a dominant starlike appearance,
and  broad emission  lines.   2MASS quasars  represent  a much  redder
near-IR-to-optical quasar  population compared to PG  quasars but have
similar $K_{s}$-band  luminosity \citep{Smith02}.  Unlike  PG quasars,
the 2MASS  and 3CR samples  include objects with  narrow, intermediate
and broad emission lines.

\begin{figure}
\epsscale{1.2}
\plotone{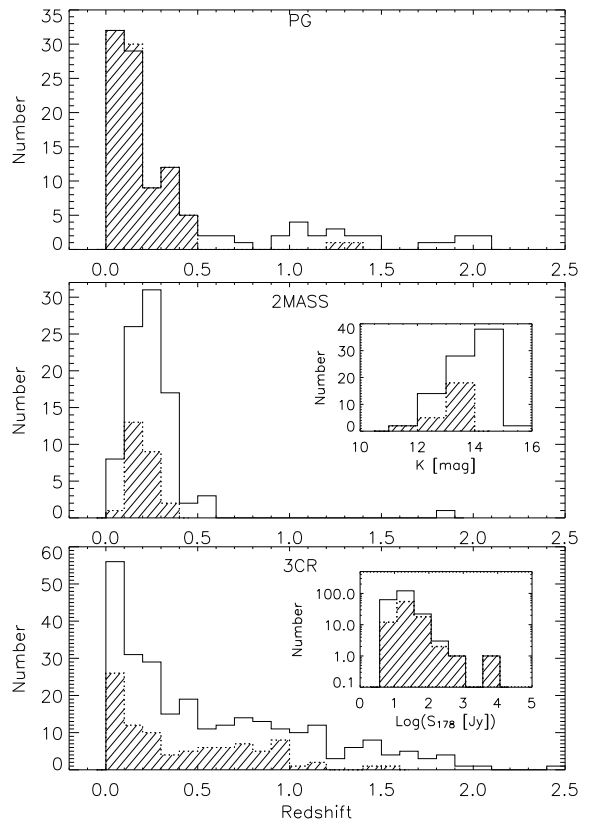}
\caption{
\label{Archive_Complete} The redshift distribution of the three subsamples 
in  this study  (shaded  area) compared  to  the corresponding  parent
samples for  the PG, 2MASS and 3CR objects.   The insert plots  show the
flux distribution of the subsample (shaded area) and the corresponding
parent sample for the 3CR and 2MASS objects. }
\end{figure}

Besides IRS spectra observed in our own programs (Program-ID 49, PI F.
Low; Program-ID 82,  PI G.  Rieke; Program-ID 3624,  PI R.  Antonucci;
Program-ID 20142,  PI P. Ogle),  we searched for archived  spectra for
objects  in  the  three  parent  samples.  Our  sample  is  listed  in
Table~\ref{Quasar_PAH}.    Fig.~\ref{Archive_Complete}   compares  the
final three  subsamples with  their corresponding parent  samples. For
the PG  parent sample from \citet{Schmidt83}, we  exclude a non-quasar
object PG 0119+229 and correct the redshift of PG 1352+011 to be 1.121
according      to       \citet{Boroson92}.       As      shown      in
Fig.~\ref{Archive_Complete},  we  have included  the  whole PG  parent
sample  at $z<$0.5.  The quasar  PG 2349-014  is not  included  in the
original PG  parent sample and  this is why  our PG subsample  has one
more  object  in the  second  redshift bin.   For  the  2MASS and  3CR
subsamples at  $z<$0.5 and z$<$1.0,  respectively, about one  third of
the objects  are included in this  study.  The subplots  show that our
2MASS and 3CR subsamples are  strongly biased toward high flux density
at the wavelength where their parent samples are selected.


\subsection{Data Reduction}

The  spectra were  obtained with  the  IRS using  the standard  staring
mode. The  intermediate products of  the {\it Spitzer}  Science Center
(SSC) pipeline  S13.0.1, S13.2.0 and S15.3.0 were processed within the
SMART software  package \citep{Higdon04}.  For  a detailed description
of  the   data  reduction,  see   \citet{Shi06},  \citet{Hines06}  and
\citet{Bouwman06}.

The slit  widths of the short-low  (SL) and long-low  (LL) modules are
3$\farcs$6  and 10$\farcs$5,  respectively.  In  order to  measure the
star  formation  from the  entire  galaxy,  we  need to  evaluate  the
extended IR emission outside of the  IRS SL slit. The SL slit width is
several hundreds  of parcsecs for  3C 272.1 and  3C 274, 2-10  kpc for
sixty-one   objects  ($z<$0.17)   and  $>$10kpc   for   the  remaining
objects. For 3C 272.1, the  MIPS image shows extended IR emission from
the  host  galaxy and  that  this emission  is  thermal  based on  the
extrapolation from radio data.  The  extended IR emission of 3C 274 is
dominated by non-thermal emission  \citep{Shi07} and is not related to
star formation. For objects  with physical slit widths between $\sim$2
kpc and 10 kpc, a total  of seventeen objects show excess IR fluxes in
the  LL  modules  compared  to  the SL  modules.   However,  the  flux
difference between  the SL and LL  modules can be  caused by different
slit-loss  due to  pointing  errors, not  necessarily  by extended  IR
emission outside the SL module slit.   For 14 out of these 17 objects,
we obtained archived  MIPS 24 $\mu$m images and  measured the FWHMs of
the radial brightness profiles.  All of them show FWHMs smaller than 3
pixels (the PSF has a FWHM of 2.4 pixels), implying that the excess IR
fluxes in the LL modules are  not due to extended IR emission from the
host galaxies.  For the remaining three objects without MIPS 24 $\mu$m
images, we use 2MASS K-band images and find that the excess flux of LL
relative to SL for one object  (PG 2304+042) may be due to extended IR
emission.  For objects with slit  widths larger than 10 kpc, we simply
assume that the IRS slit contains  all the IR emission from the galaxy
and that the mismatch between the SL and LL spectra is due to variable
slit-loss. Therefore, except  for 3C 272.1, 3C274 and  PG 2304+042, we
rescale the  SL spectra so  that the SL  and LL spectra have  the same
flux density at 14.2 $\mu$m.

\subsection{The Extraction of Aromatic Features}

\begin{figure}
\epsscale{1.2}
\plotone{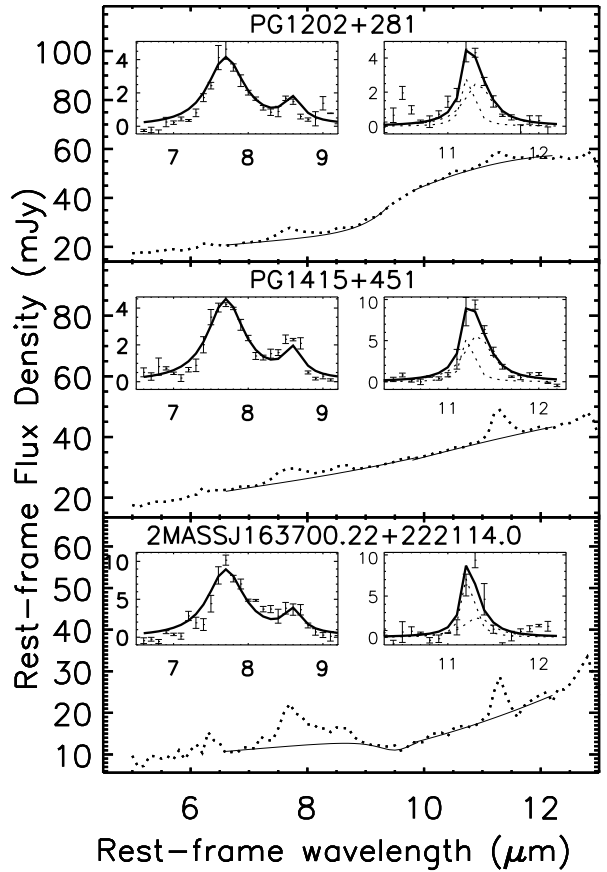}
\caption{ \label{Spec_example} 
Examples of  the extraction of the 7.7  and 11.3 um aromatic 
features in  the spectra with  silicate emission, no  silicate feature
and  silicate absorption,  respectively. The  dotted lines  are  the IRS
spectra while  the solid  lines are the  continua.   The  subplots  show the  Drude
profiles of the two features where the 11.3 um feature is fitted with
two Drude profiles (dotted lines). }
\end{figure}

\setcounter{figure}{2}
\begin{figure*}
\epsscale{1.2}
\plotone{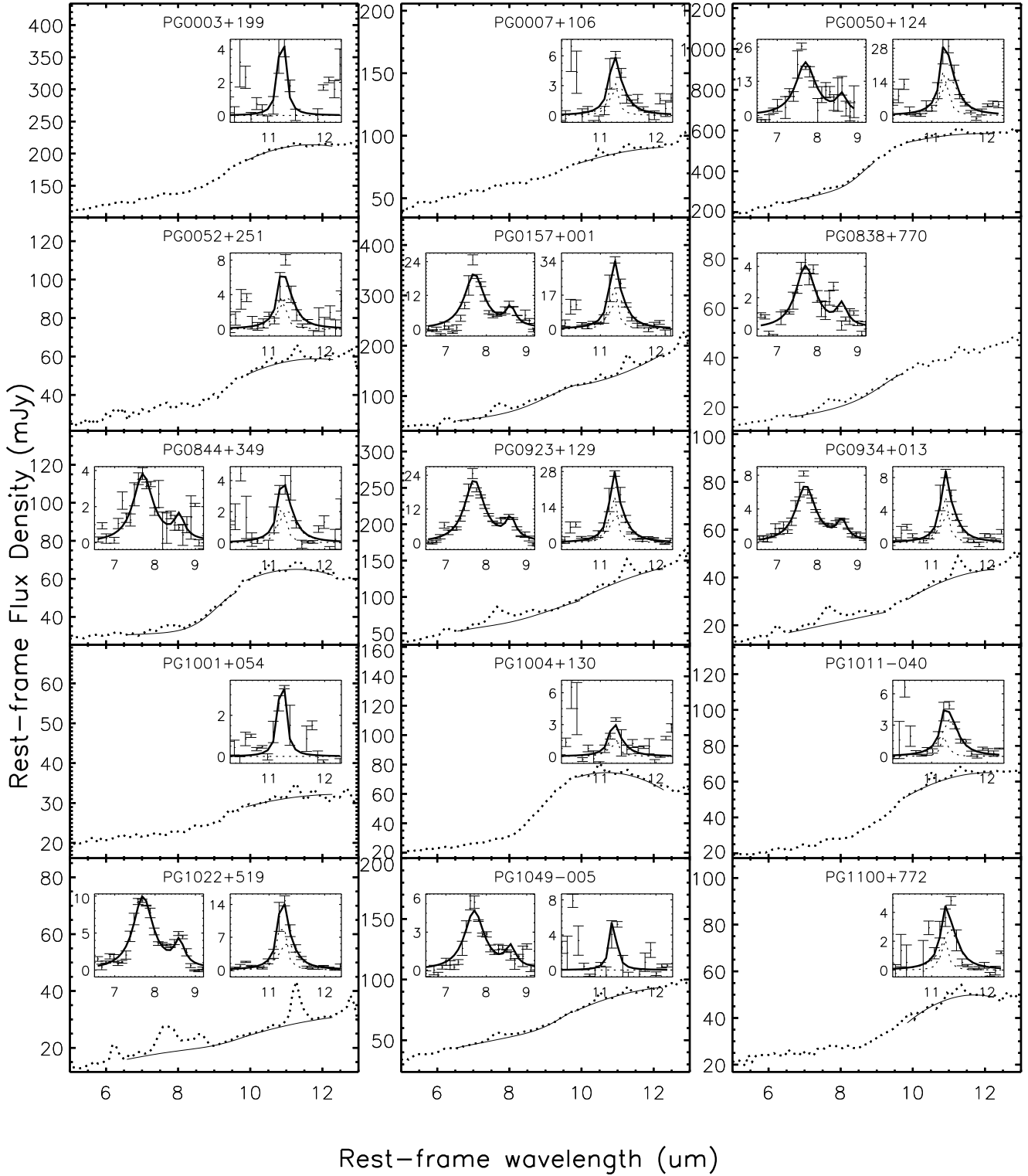}
\caption{\label{spectra_PAH} 
IRS spectra of AGN with detected aromatic features.  The solid line
is  the derived continuum  for  the  7.7 $\mu$m  and/or  11.3 $\mu$m  aromatic
features. The subplots show the Drude profiles of the two features. }
\end{figure*}

\setcounter{figure}{2}
\begin{figure*}
\epsscale{1.20}
\plotone{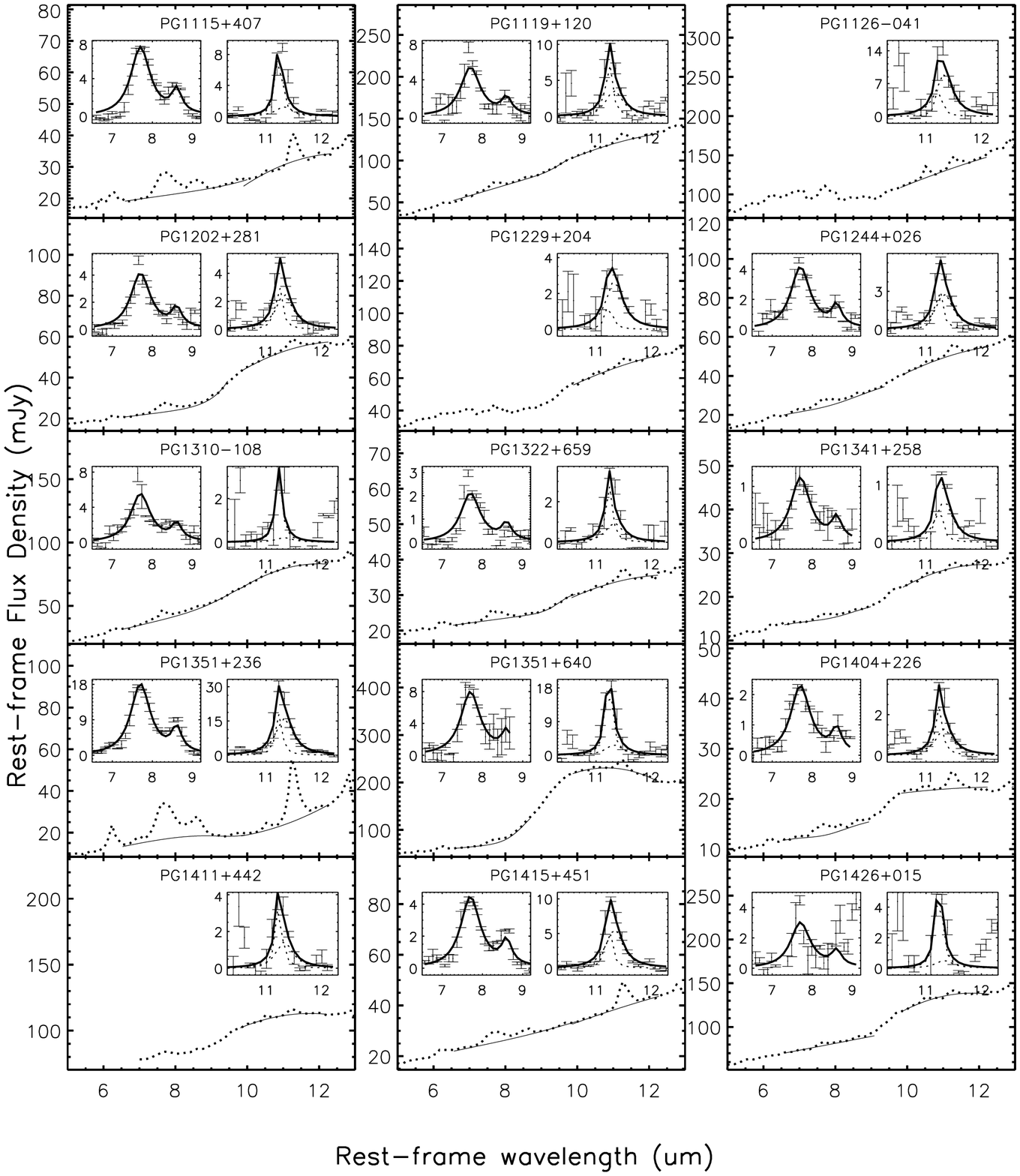}
\caption{ Continued.}
\end{figure*}

\setcounter{figure}{2}
\begin{figure*}
\epsscale{1.2}
\plotone{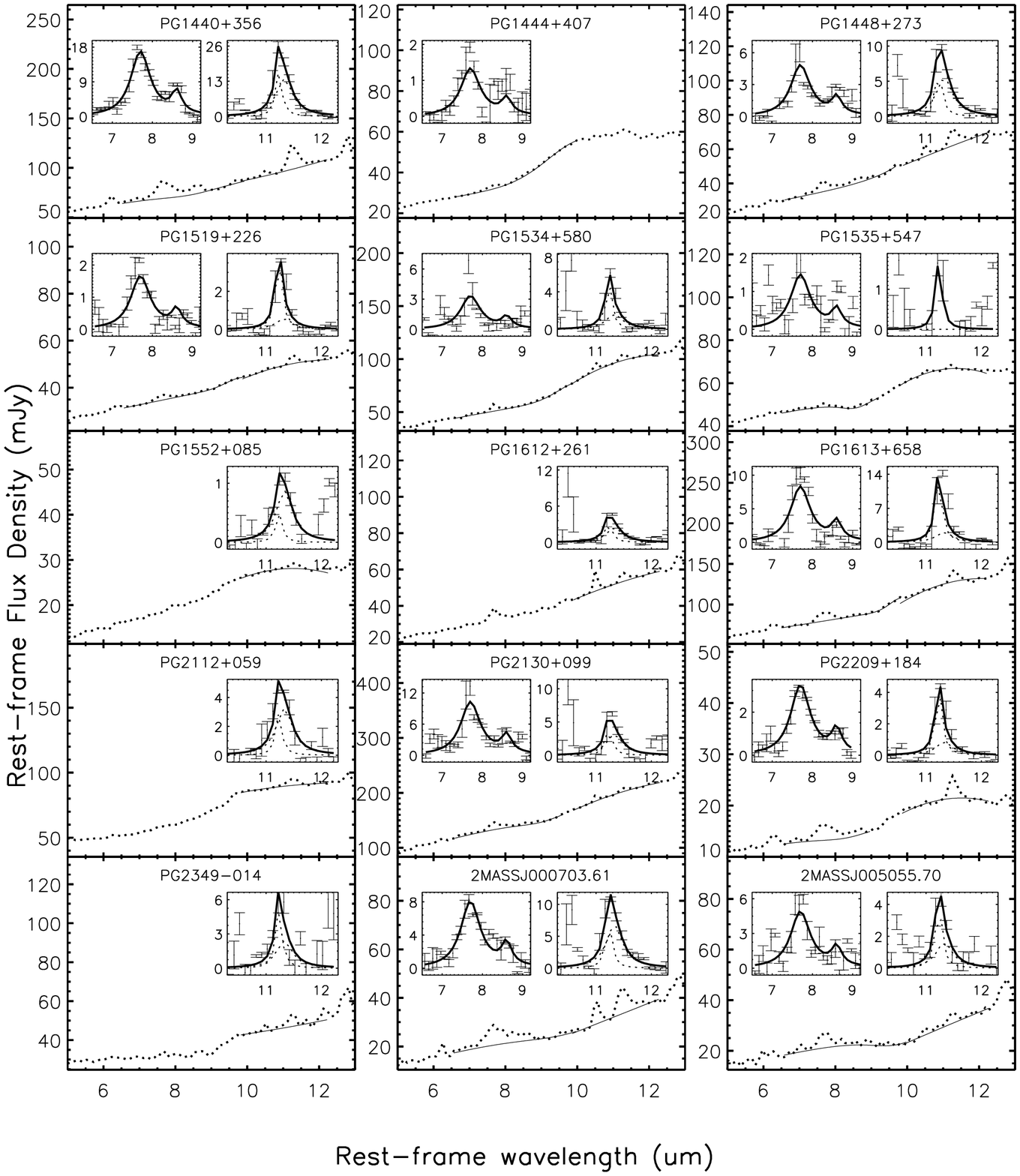}
\caption{ Continued.}
\end{figure*}

\setcounter{figure}{2}
\begin{figure*}
\epsscale{1.2}
\plotone{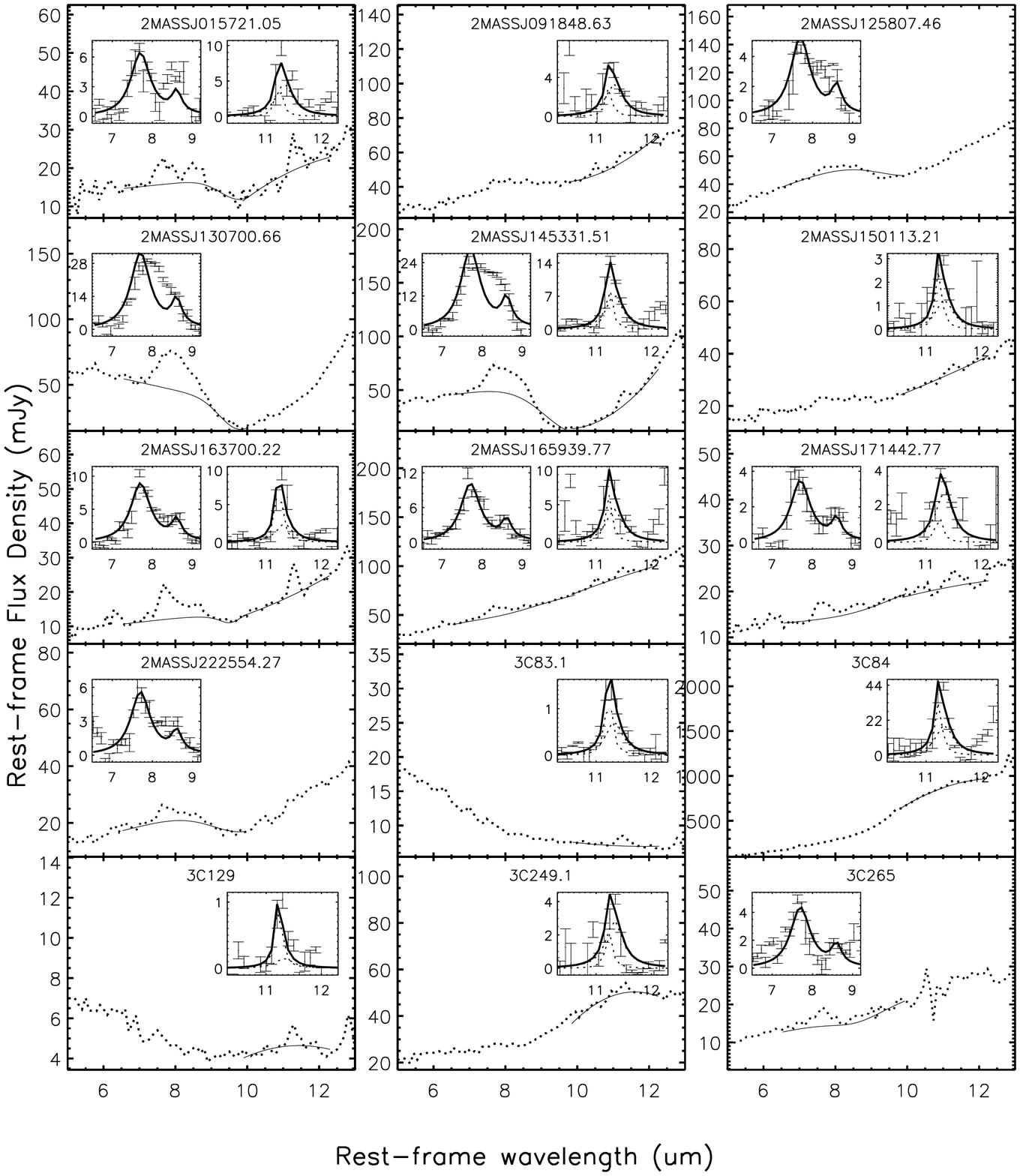}
\caption{ Continued.}
\end{figure*}

\setcounter{figure}{2}
\begin{figure*}
\epsscale{1.2}
\plotone{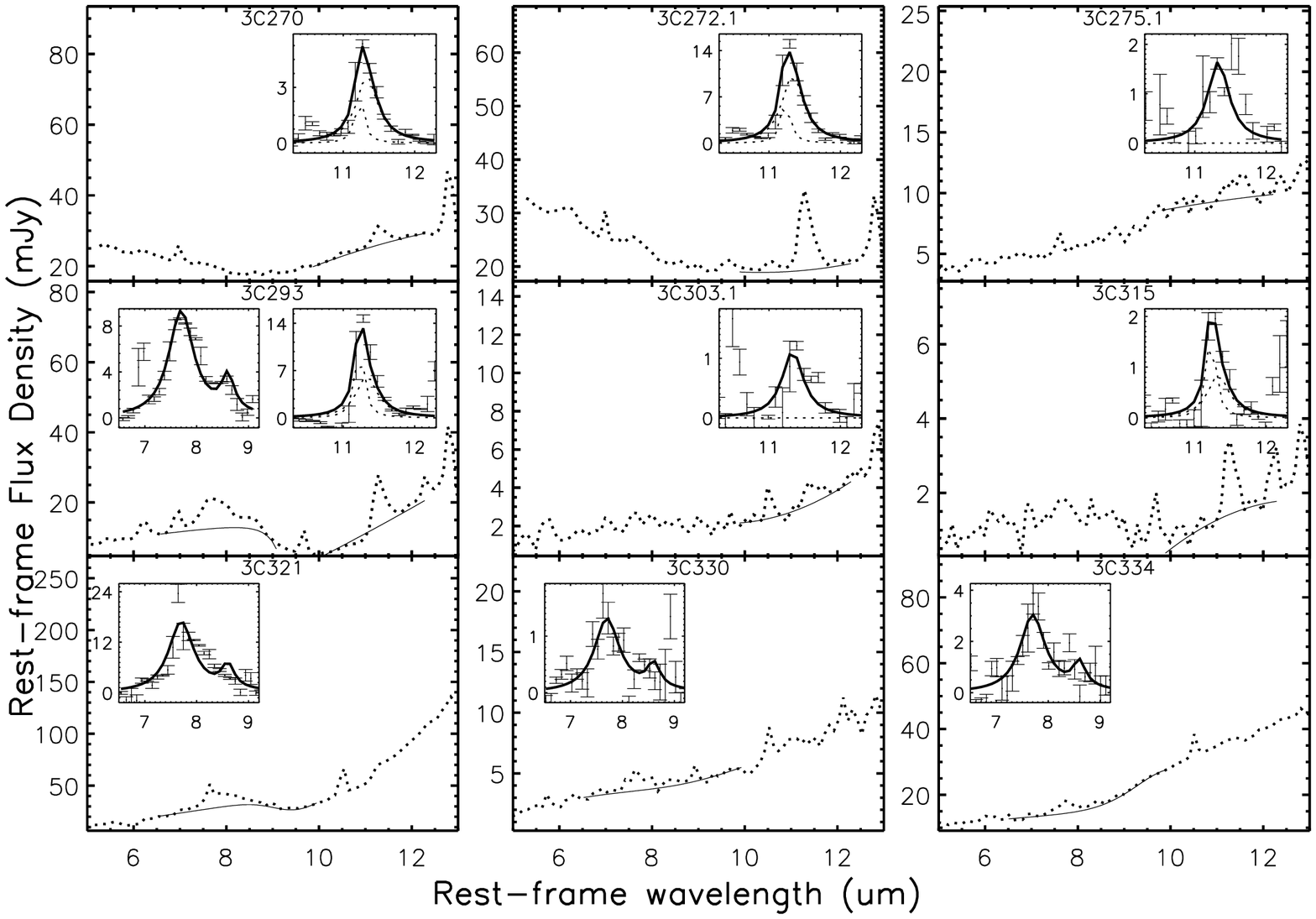}
\caption{Continued.}
\end{figure*}

The  7.7  $\mu$m feature  resides  at the  blue  end  of the  silicate
feature. The  level of  contamination by the  silicate feature  on the
aromatic flux  measurement depends  on several factors,  including the
strength of the  silicate feature, the shape of its  blue wing and the
shortest  wavelength that  the  blue  wing extends  to.   As shown  in
\citet{Hao05} or our  Figure~\ref{spectra_PAH}, all these factors vary
in different  sources, resulting in  deviations from the  line profile
for  a  normal  galaxy  interstellar  medium.  To  account  for  these
variations,  we fit  the  blue wing  of  the silicate  feature with  a
Doppler profile:
\begin{equation}
f_{\lambda}=\frac{f_{\lambda_{0}}}  {  (\lambda  - \lambda_{0})^{2}  +
(\alpha_{L})^{2}}exp( -( (\lambda - \lambda_{0})/\alpha_{D} )^{2} )
\end{equation}
where $\lambda_{0}$  can be interpreted  as the central  wavelength of
the  silicate  feature,  and   the  combination  of  $\alpha_{D}$  and
$\alpha_{L}$  control the  shape of  the  blue wing  and the  starting
wavelength  where the  silicate feature  arises.  The  profile  has no
physical meaning and is adopted  only for practical purposes. As shown
in Figure~\ref{Spec_example}, it can fit the 7.7 $\mu$m feature well.

The  procedure  to extract  the  7.7  $\mu$m  aromatic feature  is  as
follows.  The spectra are first rebinned to a resolution of 0.1 $\mu$m
to  remove multiple  points at  the same  wavelength, using  the SMART
software.  The  continua underlying  the 7.7 $\mu$m  aromatic features
and silicate  features are defined  as power laws over  three spectral
windows,  5.2-5.5 $\mu$m,  5.5-5.8 $\mu$m  and 6.7-7.0  $\mu$m.  These
spectral regions are selected to avoid the possible ice feature at 6.0
$\mu$m  and  aromatic  features  at  6.2  $\mu$m.   We  then  fit  the
continua-subtracted spectra simultaneously  with two aromatic features
at 7.7  and 8.6 $\mu$m  and the silicate  feature.  The shapes  of the
aromatic features are assumed to be Drude profiles.  Due to the low EW
of aromatic  features in  AGNs, the  FWHMs of the  7.7 and  8.6 $\mu$m
features are fixed at 0.6  and 0.3 $\mu$m, respectively. The height of
the 8.6  $\mu$m feature is also fixed  to be one-third of  that of the
7.7 $\mu$m feature.   This relative height is similar  to those of two
average   spectra   of    HII-like   nearby   galaxies   obtained   by
\citet{Smith07}.  For the silicate feature,  we fit only the blue wing
with a Doppler profile.  The starting wavelength of the spectral range
for the fit is  fixed at 6.5$\mu$m.  We vary the red  end from 9 to 12
$\mu$m  to have the best  fit judged  by visual  inspection.  For  most of
the sources, the measured aromatic flux depends little on the selected red
end wavelength.  The  feature is considered detected if  the height of
the 7.7  $\mu$m feature is five  times greater than the  mean noise in
the continuum.

For  the 11.3  $\mu$m feature,  the  silicate feature  behaves like  a
continuum  and the  slope of  the underlying  silicate  profile varies
smoothly  across the  aromatic  feature.  Therefore,  we  are able  to
determine the silicate profile  simply with a quadratic interpolation.
The 11.3 $\mu$m feature is  fitted with two Drude profiles centered at
11.23 and  11.33 $\mu$m  with fixed FWHMs  of 0.135 and  0.363 $\mu$m,
respectively. The  combination of these  two Drude profiles  fits well
the 11.3 $\mu$m features of nearby galaxies, as demonstrated with high
S/N IRS spectra by \citet{Smith07}.  After the spectrum is rebinned to
a  resolution of  0.1 $\mu$m,  the continuum  (plus  silicate feature)
shape  is defined  by using  a quadratic  interpolation over  the four
continuum  spectral  regions,  9.7-10.0, 10.0-10.3  $\mu$m,  10.7-11.0
$\mu$m and  11.7-12.1 $\mu$m.   We then fix  the continuum  shape, the
FWHM and the  center wavelength of the two  Drude profiles, but adjust
the normalization of the continuum  and the strength of Drude profiles
to  fit the  spectra  in the  range  including the  continuum and  the
feature  (9.7-10.3  $\mu$m  and  10.7-12.1 $\mu$m).   The  feature  is
considered detected if the height  of the combination of the two Drude
profiles is five  times greater than the mean  noise in the continuum.
If  the feature  is not  detected, the  upper limit  is  calculated by
assuming the same relative strength of the two Drude profiles as given
by the fit  and taking five times the mean noise  for the total height
of the two profiles.  We  visually inspected each detected feature and
found that the 11.3 $\mu$m features  of eleven objects may not be real
due to larger  noise around the feature relative to  the mean noise in
the  continuum.  For fifteen  objects, the  continuum was  also fitted
with an alternative quadratic interpolation,  due to a large change in
the  slope of  the silicate  profile around  the 11.3  $\mu$m feature.
However,  the difference  in the  feature strength  is smaller  than a
factor of 1.5,  showing that the continuum fitting  procedure does not
affect our results strongly.

To  test the  robustness of  our procedures  against  strong continua,
power-law  continua   with  different  strengths  are   added  to  the
star-forming  templates from  \citet{Dale01} and  \citet{Dale02}.  The
7.7 and  11.3 $\mu$m aromatic  features are extracted using  the above
procedures and  the flux  variations are smaller  than 1\% for  the EW
range from the original value ($\sim$1$\mu$m) to 0.01 $\mu$m.

\subsection{Uncertainty of the Aromatic Flux}

We have evaluated each step in extracting the features to estimate the
final uncertainty  of the aromatic  flux. To estimate  the uncertainty
due to  the rebinned spectral resolution, the  fluxes are re-measured
 with  rebinned resolutions from  0.08   to 0.12
$\mu$m   for  features   observed  with   SL  module   (resolution  of
$\sim$0.1$\mu$m).  For the objects  at z$>$0.24, where the 11.3 $\mu$m
feature is observed with  LL module ( resolution of $\sim$0.28$\mu$m),
we compare the measured flux for rebinned resolutions ranging from 0.1
to  0.3  $\mu$m. Comparing  these  measurements  to  the feature  flux
obtained  at a rebinned  resolution of  0.1 $\mu$m,  we find  that the
differences are always below 10\%.

To estimate the uncertainty caused by  the photon noise and the fit of
the continuum  and silicate feature,  we produce a  noiseless spectrum
for each detected aromatic  feature.  The simulated noiseless spectrum
for the  7.7 $\mu$m feature  is the measured power-law  continuum plus
the measured  Dopper profile  of the silicate  feature plus  two Drude
profiles of  the measured 7.7  and 8.6 $\mu$m features.   The spectrum
for  11.3  $\mu$m  is  the quadratically  interpolated  continuum  and
silicate  profile plus two  measured Drude  profiles. We  then perturb
this noiseless spectrum  100 times to produce noisy  spectra with mean
S/N  equal  to  the  observed  S/N. The  aromatic  features  are  then
extracted  from  these simulated  spectra  in  the  same way  and  the
1-$\sigma$  uncertainty  is obtained  as  the  difference between  the
original flux  and those from the simulated  spectra.  The uncertainty
in  this step is  typically $<$15\%  for the  11.3 $\mu$m  feature and
$<$30\% for the 7.7 $\mu$m feature.

Due to the contamination by the silicate feature, we are unable to fit
the 7.7 $\mu$m  feature with multiple Drude profiles.   To compute the
uncertainty  in the  assumed profile  with a  fixed FWHM  for  the 7.7
$\mu$m  feature,  we  have  used  the  code  (PAHFIT.pro)  written  by
\citet{Smith07} to measure accurate fluxes for the 7.7 $\mu$m aromatic
complexes  of  the  four  composite  spectra  of  nearby  galaxies  in
\citet{Smith07}.  We then re-construct the 7.7 $\mu$m profile with the
fitted parameters  and measure  the flux with  a single  Drude profile
with a FWHM of 0.6 $\mu$m.  The difference in fitted feature strengths
is around  10\%, which is  adopted as the  uncertainty due to  the 7.7
$\mu$m aromatic  profile.  No uncertainty  is applied for  the assumed
profile of the 11.3 $\mu$m feature.  The above uncertainties are added
quadratically to give  the final error of the  measured aromatic flux.
Table~\ref{Quasar_PAH}  lists the  measured fluxes,  uncertainties and
EWs for both aromatic features.

\section{EXCITATION MECHANISM OF AROMATIC FEATURES IN AGNS}

As shown in \S~1, the low  EW of the aromatic features and the spatial
extension  of the aromatic  emission in  active galaxies  suggest that
these  features   are  most  likely  predominantly   excited  by  star
formation.   With the  significant  number of  detections of  aromatic
features in this study, we can test this hypothesis.

\subsection{The Profile of Aromatic Features in AGN}

\subsubsection{The Composite Spectra}

\begin{figure}
\epsscale{1.2}
\plotone{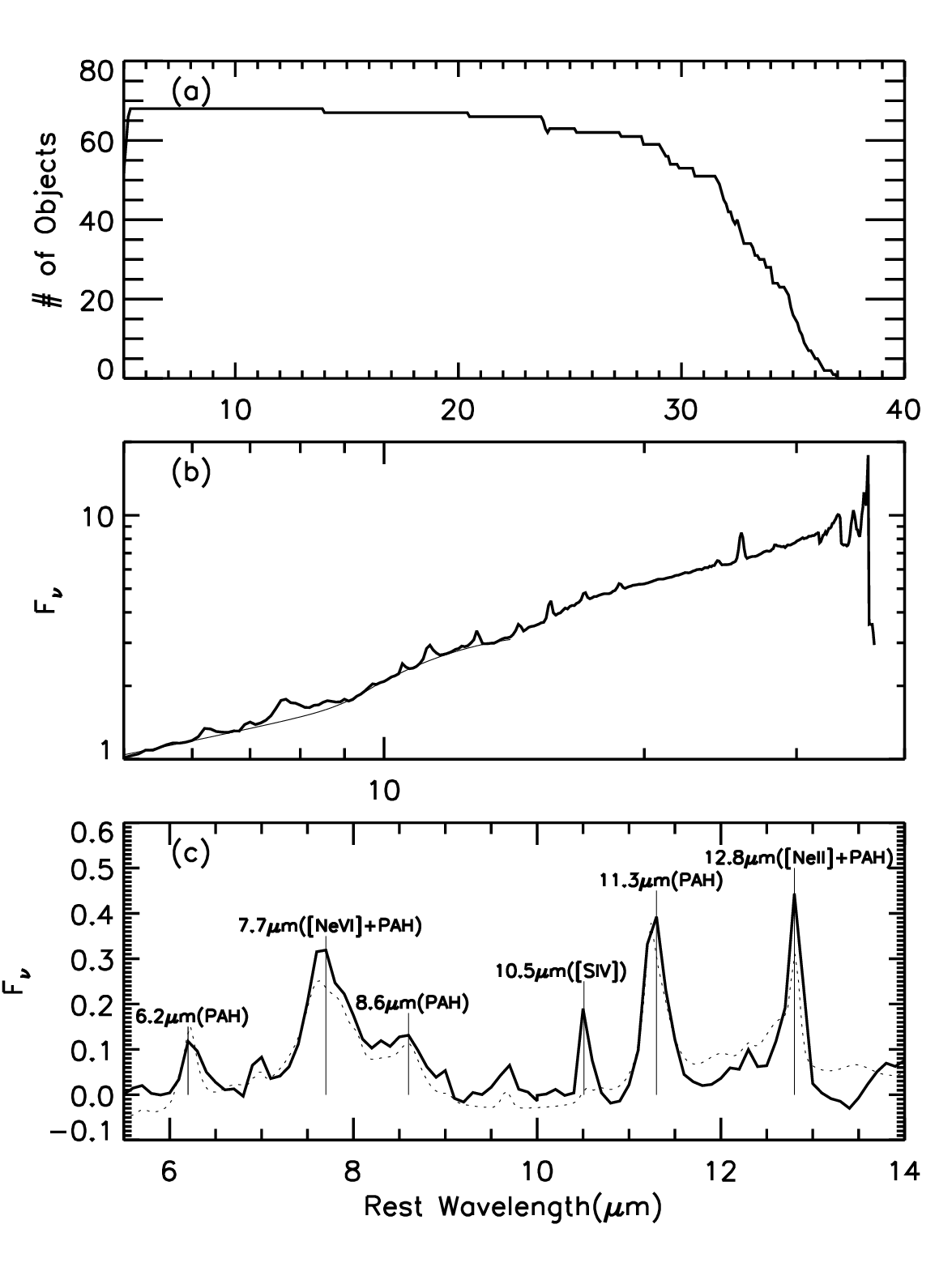}
\caption{ \label{CP_arith_mean_Dec} 
(a)  The number of  objects in  each wavelength  bin of  the composite
spectrum. (b) The arithmetic mean  spectrum (the heavy solid line) of
AGN with one of the 7.7 and 11.3 $\mu$m aromatic  features detected and the
fitted continuum (the light  solid line). (c) The continuum-subtracted
spectrum (the heavy solid line) superposed with the composite spectrum
(the dotted line) of the HII-like galaxies from \citet{Smith07}. }
\end{figure}

\begin{figure}
\epsscale{1.2}
\plotone{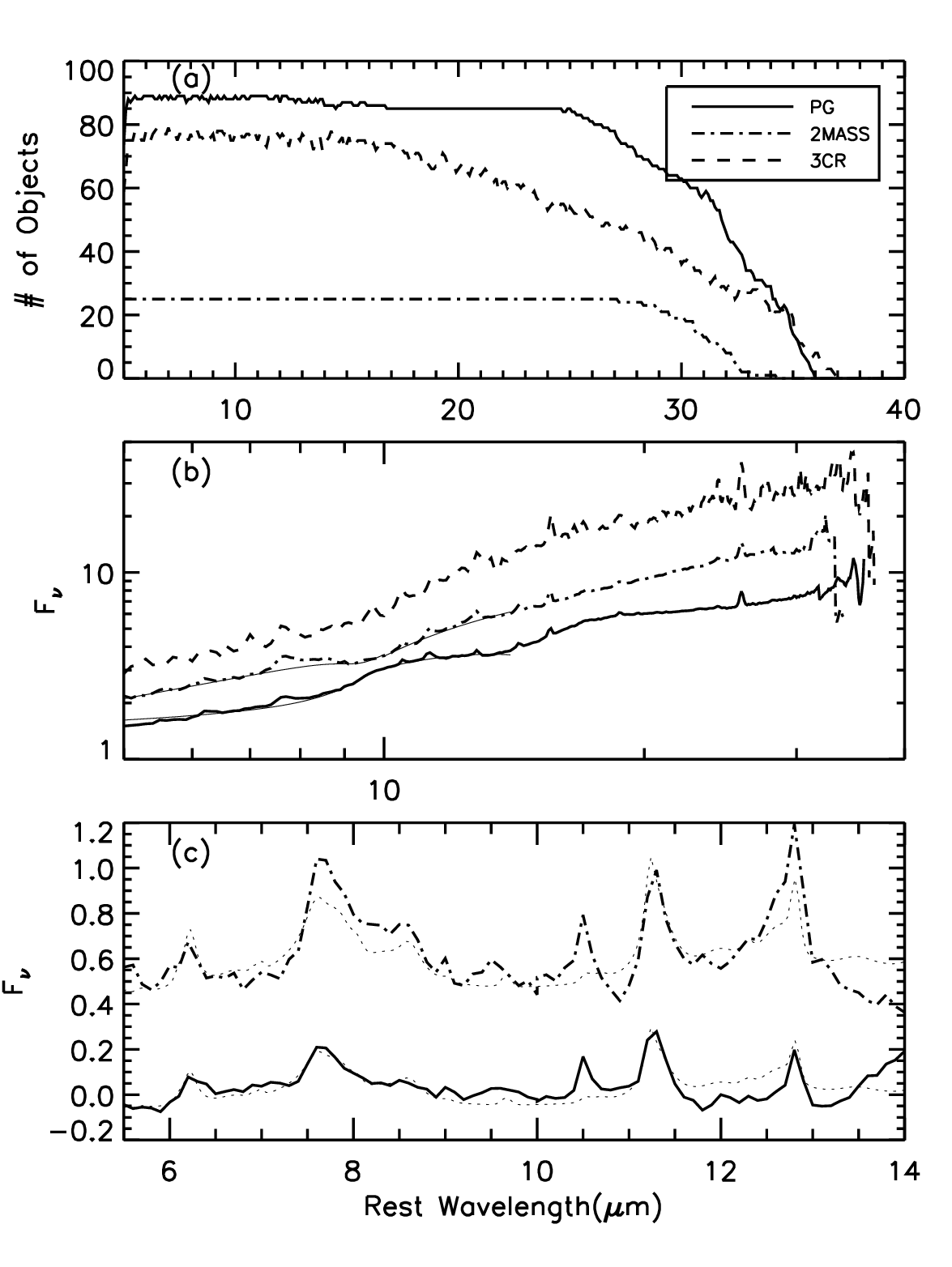}
\caption{ \label{CP_arith_mean_Type} 
(a)  The number of  objects in  each wavelength  bin of  the composite
spectra of PG, 2MASS and 3CR AGN, respectively. (b) The arithmetic mean spectra and
the   fitted    continua   (the   light   solid    lines).   (c)   The
continuum-subtracted spectra of PG  and 2MASS AGN, superposed with the
composite  spectra (the dotted  lines) of  the HII-like  galaxies from
\citet{Smith07}.}
\end{figure}

To study the profile of the aromatic features in AGN, we have produced
the composite  spectra for several  groups of objects.   The composite
spectrum   is   computed   following   the  procedure   described   in
\citet{VandenBerk01}.   All the  observed spectra  are shifted  to the
rest-frame  and   then  rebinned  to  a   common  spectral  resolution
(0.1$\mu$m)  within SMART.  After they  are ordered  by  redshift, the
first  spectrum  is   rescaled  randomly.   The  following  individual
spectrum is  rescaled to have the  same mean flux density  in a common
spectral region of the mean spectra of all lower redshift spectra. The
common spectral region is defined  to be 5.0-6.0 $\mu$m where there is
little influence  from the silicate  or aromatic features.   The final
composite  spectrum is the  arithmetic mean  of all  rescaled spectra.
Unlike  in  \citet{VandenBerk01},  we  have not  produced  the  median
spectrum since the  average one shows much higher  S/N.  As implied by
the  compositing  procedure,   the  aromatic  features  of  individual
observed spectra with  higher EW have larger weight  in the feature of
the final composite spectrum.

The first arithmetic mean spectrum is the one of AGN with at least one
of     the    7.7     and    11.3     $\mu$m     features    detected.
Fig.~\ref{CP_arith_mean_Dec}(a)  plots the number  of objects  used in
each wavelength bin.  As shown in Fig.~\ref{CP_arith_mean_Dec}(b), the
overall spectrum  shows a power-law-like continuum  with weak silicate
features.   We determined the  continuum between  5.0 and  10.0 $\mu$m
using the procedure  for extracting the 7.7 $\mu$m  feature but do not
constrain  the  strength of  the  8.6  $\mu$m  aromatic feature.   The
continuum between  9.5 and  14.0 $\mu$m is  defined to be  a quadratic
interpolation over  the mean flux  densities of four  spectral regions
(10.0-10.3, 10.8-11.0, 13.0-13.2, and  13.4-13.6 $\mu$m).  As shown in
Fig.~\ref{CP_arith_mean_Dec}(c),  broad features  are  present at  6.2
$\mu$m, 7.7 $\mu$m,  8.6 $\mu$m, 11.3 $\mu$m and  12.8 $\mu$m, similar
to those  in star forming galaxies  \citep[See][]{Lu03, Smith07}.  The
dotted  curve   in  Fig.~\ref{CP_arith_mean_Dec}(c)  shows   the  mean
spectrum  of   two  composite   spectra  of  HII-like   galaxies  from
\citet{Smith07} where  the spectrum is  shifted and rescaled  to match
the 11.3  $\mu$m feature of the  AGN spectrum.  There is  only a small
discrepancy  in the  shapes  and relative  strengths  of the  aromatic
features between AGN and HII-like  galaxies.  A small amount of excess
emission at 7.7 and 12.8 $\mu$m in the AGN spectrum is most likely due
to [NeV]7.65$\mu$m  and [NeII]12.8$\mu$m, respectively,  as the excess
emission  shows a  narrow  FWHM.  The  result  indicates the  observed
aromatic features in AGN resemble those in star-forming galaxies.  The
composite spectrum  of AGN without either feature  detected still does
not show detectable aromatic features.

Fig.~\ref{CP_arith_mean_Type} shows  the arithmetic mean  spectra for
PG,  2MASS  and  3CR  objects, respectively.   The  silicate  emission
features  are present  in  the PG  spectrum  while the  2MASS and  3CR
spectra have  silicate absorptions.  Aromatic features  are visible in
the PG and 2MASS composite spectra, but  not in the 3CR spectrum.  As shown in
Fig.~\ref{CP_arith_mean_Type}(c),  the  comparison  to  the  HII-like
galaxies  indicates the 11.3/7.7$\mu$m  feature ratio  ($\sim$0.30) of
the PG spectrum is a little higher while the 2MASS spectrum presents a
lower ratio ($\sim$0.22). However,  they are within the one-$\sigma$ range
for star-forming galaxies as shown below.

\subsubsection{The Distribution of the Aromatic Feature Ratio}

\begin{figure}
\epsscale{1.2}
\plotone{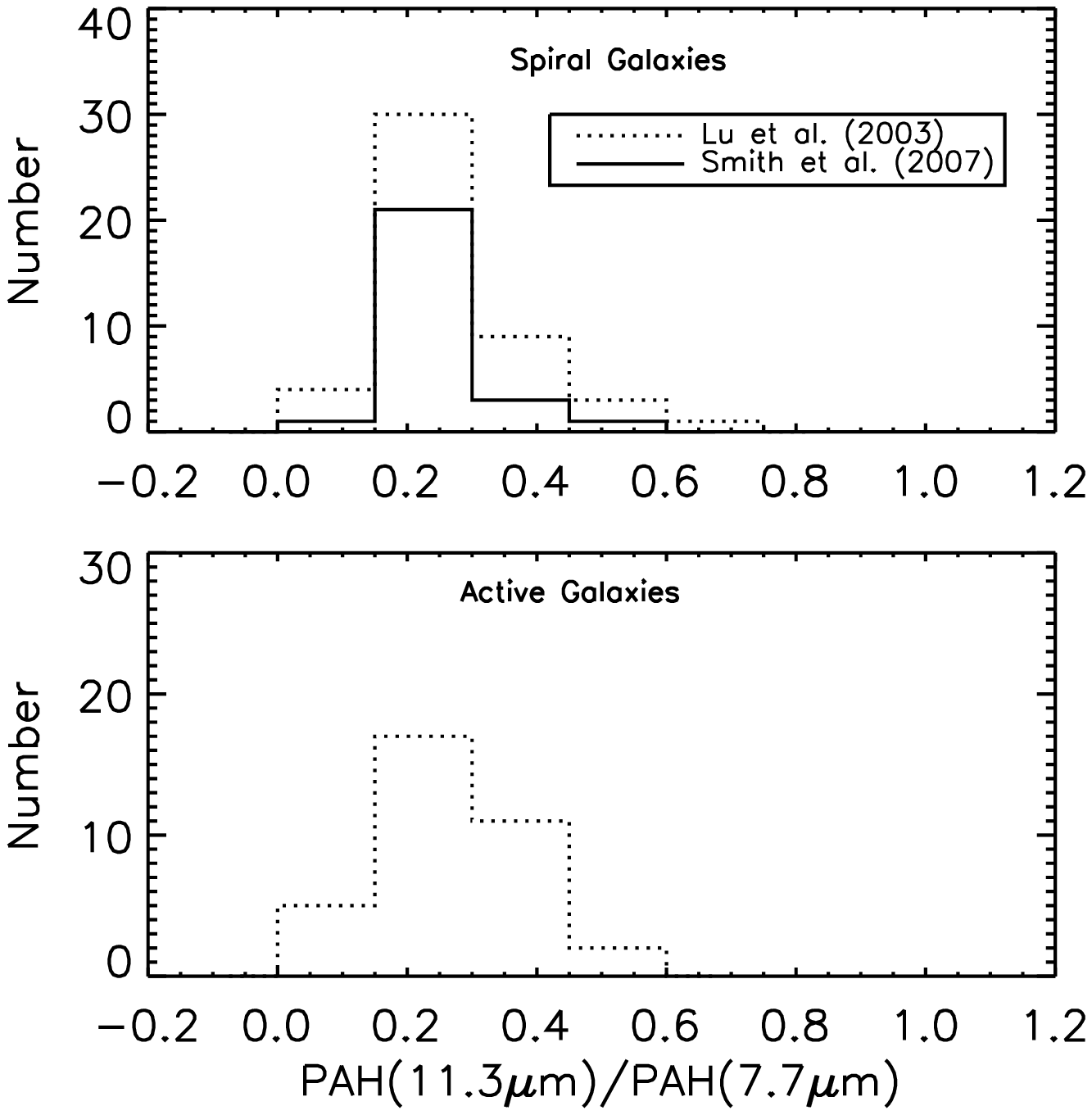}
\caption{ \label{PAH_ratio} 
The ratio  of the  11.3 $\mu$m aromatic  flux to the  7.7 $\mu$m flux.  
The upper plot  is the ratio  for normal
spiral galaxies  from \citet{Lu03} and \citet{Smith07}  while the lower  
plot is  for active galaxies in this paper.}
\end{figure}

The above  comparisons reveal that the shapes  and relative strengths
of the aromatic  features of the AGN composite  spectra are similar to
those  of   HII-like  galaxies.   Fig.~\ref{PAH_ratio}   compares  the
distribution  of the  11.3/7.7$\mu$m aromatic  ratios between  AGN and
normal  star-forming galaxies  from \citet{Smith07}  and \citet{Lu03}.
For the  sample of \citet{Smith07}, we only  include HII-like galaxies
but exclude  a low-metallicity dwarf galaxy (HoII).   No correction is
applied  to  their  aromatic  fluxes,  since they  are  obtained  with
multiple Drude  profile fitting. The flux  of the 7.7  and 11.3 $\mu$m
aromatic  features  quoted in  \citet{Lu03}  is  the integrated  value
without continuum subtraction from 7.20  to 8.22 $\mu$m and from 10.86
to 11.40 $\mu$m, respectively.  We correct their ratios by a factor of
1.08 to account  for the difference between their  measured fluxes and
the Drude-profile fluxes  used in this paper. This  factor is obtained
based  on   the  four  composite  spectra  of   nearby  galaxies  from
\citet{Smith07}.  In  the \citet{Lu03} sample, one  object is excluded
since  the  integrated aromatic  flux  contains  significant hot  dust
emission.

As  shown in  Fig.~\ref{PAH_ratio}, the  flux ratio  of AGN  with both
features detected  has a similar  distribution to that  of star-forming
galaxies.    The  mean   11.3/7.7-aromatic  ratio   for  the   AGN  is
0.27$\pm$0.1,  compared with 0.28$\pm$0.11  and 0.26$\pm$0.07  for the
spiral  galaxies of  \citet{Lu03}  and \citet{Smith07},  respectively.
The Kolmogorov-Smirnov (K-S) test  indicates a probability of 99\% and
40\% that  our AGN sample is  the same as the  star-forming galxies of
\citet{Lu03} and \citet{Smith07}, respectively.

Variations of the aromatic flux ratio have been observed among regions
covering   a  wide   range   of  physical   and  chemical   properties
\citep[e.g.][]{Roelfsema96, Vermeij02}. On  the other hand, studies of
the aromatic features in the same environment show that the flux ratio
is   insensitive   to   the   intensity   of   the   radiation   field
\citep{Uchida00,  Chan01}.   Among  different  galaxies, there  is  no
systematic difference in the aromatic flux ratio with the intensity of
the radiation field, as seen by \citet{Lu03} where the spiral galaxies
studied have  total IR luminosity spanning from  10$^{9}$ to 10$^{11}$
L$_{\odot}$.   This may  arise  because various  aromatic regions  are
averaged out over the entire  galaxy.  The similar distribution of the
ratio between AGN and spiral galaxies as shown in Fig.~\ref{PAH_ratio}
implies that the features observed in AGN are excited under conditions
similar  to   those  averaged  over  normal   star  forming  galaxies.
\citet{Smith07} have  found that 20\% of  galaxies with low-luminosity
active nuclei show a weak  7.7 $\mu$m feature relative to the strength
of the 11.3 $\mu$m feature.  The  origin of this deviation is not well
understood.  However, if it is the nuclear radiation that accounts for
this peculiar ratio, the similar  feature ratio between our sample and
star-forming  galaxies indicates  the aromatic  feature output  in our
sample is dominated by star  formation, not by the active nuclei.  For
objects with only  one detected feature, the distribution of
the limits on  $F_{7.7{\mu}m}/F_{11.3{\mu}m}$ is still consistent with
that of star-forming galaxies.

\subsection{The Global IR SED}

\begin{figure}
\epsscale{1.2}
\plotone{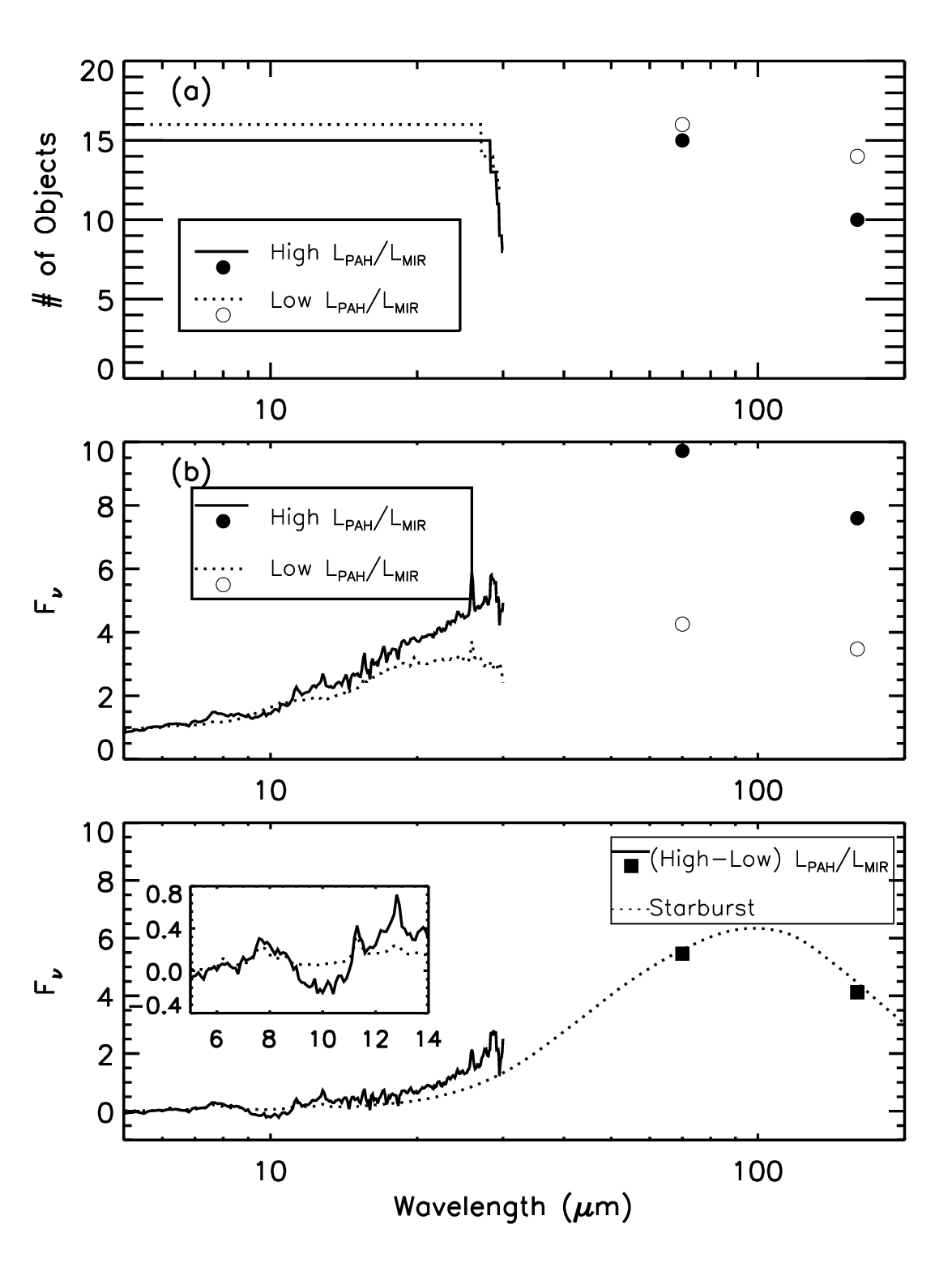}
\caption{ \label{CP_geom_mean} 
(a)  The number of  objects in  each wavelength  bin of  the composite
spectra  of  the high-$L$(PAH)/$L$(MIR)  subsample  (solid line  plus
filled circles)  and the low-$L$(PAH)/$L$(MIR)  subsample (dotted line
plus  open circles),  where $L$(MIR)  is the  total  mid-IR luminosity
between 5.0  and 6.0  $\mu$m and $L$(PAH)  is the  11.3$\mu$m aromatic
luminosity or the 7.7$\mu$m aromatic luminosity multiplied by a factor
of 0.27 for objects with only the 7.7 $\mu$m feature detected.  (b) The
geometric  mean  spectra  of  the two  subsamples.  (c)  The  spectrum  of
high-$L$(PAH)/$L$(MIR) minus low-$L$(PAH)/$L$(MIR) objects superposed on the
starburst template  with $L_{8-1000{\mu}m}$=2.0$\times$10$^{11}$ L$_{\odot}$ from
\citet{Dale01} and \citet{Dale02}. }
\end{figure}

The global IR SED of AGN is affected by many factors.  However, if the
aromatic feature  originates from star-forming  regions, the composite
spectrum of the subsample with  a higher fraction of aromatic emission
in  the  mid-IR emission  should  show  a  higher fraction  of  far-IR
emission.

Fig.~\ref{CP_geom_mean} compares  the composite spectra from  5 to 200
$\mu$m  for high-$L$(PAH)/$L$(MIR) and  low-$L$(PAH)/$L$(MIR) objects,
where  $L$(MIR) is  the total  mid-IR luminosity  between 5.0  and 6.0
$\mu$m and $L$(PAH) is the  11.3 $\mu$m aromatic luminosity or the 7.7
$\mu$m aromatic luminosity multiplied by  a factor of 0.27 for objects
with only  the 7.7  $\mu$m feature detected.   We define  the dividing
value  of  $L$(PAH)/$L$(MIR)  for  all  objects with  MIPS  70  $\mu$m
measurements  so that  the high  and  low-$L$(PAH)/$L$(MIR) subsamples
have  similar  numbers  of  objects.   The objects  with  upper  limit
measurements  for  the  aromatic  fluxes  are also  included  for  the
low-$L$(PAH)/$L$(MIR)  subsample while  only  feature-detected objects
are  included  for  the  high-$L$(PAH)/$L$(MIR)  subsample.   We  have
produced geometric  mean composite spectra, which  conserve the global
continuum shape \citep[See][]{VandenBerk01}.   For each subsample, the
IRS  spectra  are  redshifted   and  rebinned  to  a  common  spectral
resolution (0.1$\mu$m).   The MIPS fluxes are  K-corrected by assuming
$\alpha$=1  and  $\alpha$=0.0  (f$_{\nu}$ $\propto$  $\nu^{-\alpha}$),
respectively, based  on the  IR SED of  the AGN in  \citet{Haas03} and
\citet{Shi05}.   Then each  spectrum is  normalized by  the  mean flux
density in the wavelength range between 5.0 and 6.0 $\mu$m.  The final
composite spectrum is defined as $(\prod_{i}^{n}f_{\lambda, i})^{1/n}$
where $\lambda$ is  the wavelength of a wavelength bin  and $n$ is the
total number of spectra in this bin.

Fig.~\ref{CP_geom_mean}(a)  plots  the   number  of  objects  in  each
wavelength  bin.  As shown  in Fig.~\ref{CP_geom_mean}(b),  given that
the  two  composite  spectra  have  similar  weak  silicate  features,
obscuration does  not account for the  difference in the  shape of the
SEDs.  The  high-$L$(PAH)/$L$(MIR) subsample has  relatively larger IR
emission    toward     wavelengths    longer    than     15    $\mu$m.
$f$(70$\mu$m)/$f$(5-6$\mu$m) and $f$(160$\mu$m)/$f$(5-6$\mu$m) are
redder  by   a  factor  of  2.5   compared  to  the   values  for  the
low-$L$(PAH)/$L$(MIR)   subsample.   The   redder   far-IR  color   is
consistent with the star-formation  origin of the aromatic features in
these active galaxies.

The    spectrum    of    the    high-$L$(PAH)/$L$(MIR)    minus    the
low-$L$(PAH)/$L$(MIR)     composite    spectra    is     plotted    in
Fig.~\ref{CP_geom_mean}(c).  We  match  this  residual  spectrum  with
star-forming templates from \citet{Dale02}  and find that the template
with    $L_{\rm   IR}(8$-$1000{\mu}m)$=2.0$\times$10$^{11}$L$_{\odot}$
presents the  most consistent 70/160$\mu$m color.   After scaling this
template to  the 70  $\mu$m photometry of  the residual  spectrum, the
subplot  shows a  good  match for  the  7.7 and  11.3 $\mu$m  aromatic
features, although there is  some discrepancy for the [NeII]12.8$\mu$m
line.   This match  provides further  evidence for  the star-formation
origin of the aromatic features in these AGN.

\subsection{Molecular Gas}

\begin{figure}
\epsscale{1.2}
\plotone{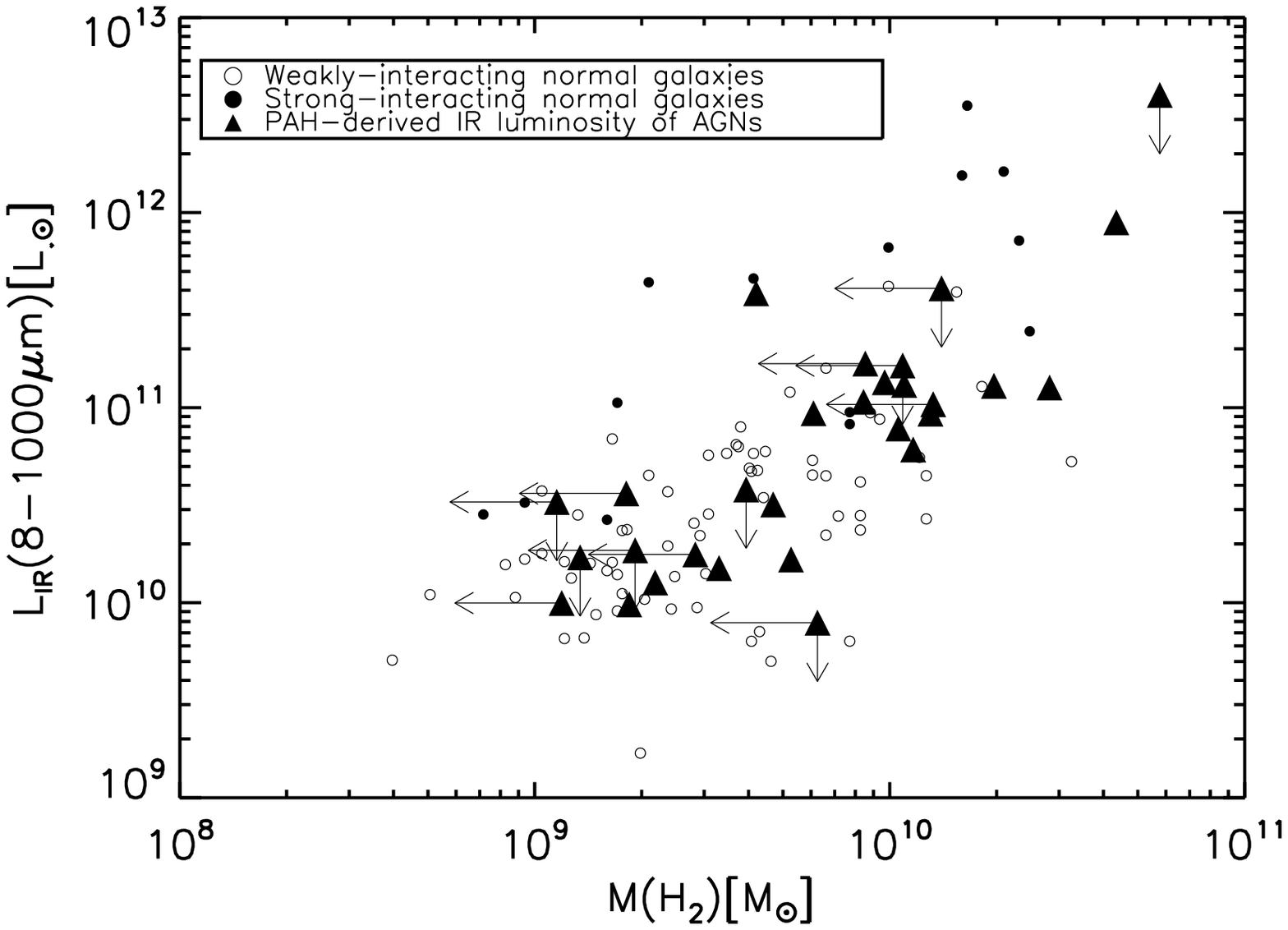}
\caption{ \label{PAH_CO} 
The plot of  the mass of CO-derived molecular  hydrogen gas versus the
aromatic-based total IR  luminosity (triangles) for AGN. Open and  filled circles
indicate weakly-interacting normal galaxies and strongly interacting normal galaxies
from \citet{Solomon88}, respectively.  }
\end{figure}

Fig.~\ref{PAH_CO} shows the mass  of CO-derived molecular hydrogen gas
versus   the   aromatic-based   star-forming  IR   (SFIR)   luminosity
(triangles).  The aromatic-based SFIR luminosity is calculated in \S~4.
The   mass    of   hydrogen   gas   is    calculated   using   $M_{\rm
H_{2}}$=1.174${\times}10^{4}$($S_{\rm           CO}$${\Delta}V$)$D_{\rm
L}^{2}$$/(1+z)$, where $S_{\rm CO}$${\Delta}V$ is the CO flux in Jy km
s$^{-1}$ and $D_{\rm  L}$ is the luminosity distance in  Mpc.  The circles
in Fig.~\ref{PAH_CO}  are the normal  galaxies from \citet{Solomon88},
where  open circles  are  for weakly-interacting  normal galaxies  and
filled circles for strongly interacting ones.  The total IR luminosity
$L_{\rm IR}$(8-1000$\mu$m) of the \citet{Solomon88} sample is computed
from   {\it  IRAS}   four-band  photometry   using  the   relation  of
\citet{Sanders96}.    The   difference   between   the   relation   of
\citet{Sanders96} and  the star-forming  templates used to  derive the
aromatic-based  SFIR  luminosity  is  typically less  than  5\%.   All
physical  parameters were  corrected to   our  adopted cosmological
model.   Fig.~\ref{PAH_CO} shows that  the behavior  of the aromatic-based
SFIR   luminosity  follows    that   of  normal   galaxies well.   The
relationship  between  the  CO   luminosity  and  SFIR  luminosity  is
consistent with the star-formation  excitation of the aromatic feature
in our AGN.

As shown above, the profile of aromatic features, the global IR SED
of AGN and  the gas content in their host  galaxies are all consistent
with  the  predominantly  star-formation  excitation of  the  aromatic
features  in  active   galaxies.  This  conclusion  confirms  previous
arguments  based  largely on  spatially  resolved  spectra of  nearby
active  gaalxies \citep[e.g.][]{Cutri84, Desert88,  Voit92, Laurent00,
LeFloch01}.

\section{The Conversion Factor from Aromatic Flux to the SFR}

Before  proceeding  with a  quantitative  study  of  the current  star
formation  around AGN  based  on  the measured  flux  of the  aromatic
features, we  need to  know how well  the aromatic features  trace the
ongoing  star-formation  activity.   For  Galactic  HII  regions,  the
variation of PAH/far-IR(40-500$\mu$m) is up to two orders of magnitude
from   ultra-compact   to    extended   optically   visible   examples
\citep{Peeters04}.  However, integrated over  the whole disk of spiral
galaxies,  the   aromatic  features  correlate   well  with  H$\alpha$
\citep{Roussel01}.   This behavior  may result  from  the galaxy-scale
quantity  averaging  out the  local  physical  properties involved  in
individual regions, such as  the escape efficiency of ionizing photons
from  HII regions  \citep[e.g.][]{Roussel01}.   The situation  becomes
complicated in the circumnuclear regions  where the EW of the observed
aromatic feature is low,  as in embedded HII regions \citep{Roussel01,
Haas02, Peeters04}. The  reason for this is unclear;  it may be caused
by obscuration, PAH  destruction, a decrease in ionizing  photons as a
result  of the  increasing  compactness  of the  HII  regions, or  the
additional  mid-IR  emission   from  highly  embedded  active  nuclei.
However, a direct attempt to  correlate the aromatic feature to far-IR
luminosity  for  star-forming galaxies  shows  that  the variation  of
PAH/far-IR is about a  factor of 2-3 \citep{Peeters04, Spoon04, Wu05}.
\citet{Spoon04}   obtained   $L$(6.2$\mu$mPAH)/$L$(IR)=0.003$\pm$0.001
from  70 normal  and starburst  galaxies.  Taking  a typical  value of
$L$(7.7$\mu$mPAH)/$L$(6.2$\mu$mPAH)=3.5      \citep{Smith07},     this
measurement               is               equivalent               to
$L$(7.7$\mu$mPAH)/$L$(IR)=0.01$\pm$0.0035.    The   aperture  mismatch
between the IR flux and the aromatic flux contributes to a part of the
scatter.                     \citet{Lutz03}                    derived
$L$(7.7$\mu$mPAH)/$L$(8-1000$\mu$m)=0.033$\pm$0.017  (assuming a Drude
profile  with 0.6  $\mu$m FWHM  for the  7.7 $\mu$m  feature)  from 10
starburst galaxies.   This ratio allows for  the aperture differences,
although the two quantities are  still not well matched.  Based on IRS
spectra of  nearby galaxies, \citet{Smith07} employed  a robust method
of  extracting aromatic  features.  The  aperture-matched  mean values
with  1-$\sigma$ uncertainties  of $L$(7.7$\mu$mPAH)/$L$(3-1100$\mu$m)
and  $L$(11.3$\mu$mPAH)/$L$(3-1100$\mu$m)  are  0.052(1$\pm$40\%)  and
0.012(1$\pm$30\%),  respectively,  for  26  HII-like  normal  galaxies
excluding one  dwarf galaxy  (Ho II) with an extreme low  ratio probably
caused by  metallicity effects  \citep[See][]{Smith07}.  A part  of the
scatter in the  ratio of $L$(PAH)/$L$(totIR) may arise  from a general
luminosity   dependence.    As   shown    in    Figure   3   of
\citet{Schweitzer06},        $L$(7.7$\mu$mPAH)/$\nu$$L_{\nu}$(60$\mu$m)
decreases     from     0.06      for     starburst     galaxies     at
$\nu$$L_{\nu}$(60$\mu$m)=1.5$\times$10$^{10}$ L$_{\odot}$ to 0.015 for
starburst-dominated   ULIRGs   at   $\nu$$L_{\nu}$(60$\mu$m)=10$^{12}$
L$_{\odot}$.
  
To  compute   the  luminosity-dependent  values,  we   have  used  the
star-forming templates  from \citet{Dale01} and  \citet{Dale02}.  Each
SED  template  is  optimized   for  a  very  narrow  luminosity  range
($\frac{{\Delta}L}{L}$  $\sim$   0.1-0.4)  where  the   luminosity  is
converted  from  the  $\alpha$  index  using  the  relation  given  by
\citet{Marcillac06}.   Aromatic  fluxes  for  all  the  templates  are
measured using  the same  procedures as for  AGN.  As  demonstrated in
\S~2.3, the  aromatic fluxes obtained  by our procedure do  not change
with the  EW, implying that there  is no systematic  difference in the
measurements of  the aromatic  fluxes between the  star-forming templates
and AGN.  The conversion factor for the 7.7 $\mu$m feature varies from
0.041  at a  SFIR luminosity  of 10$^{9}$  L$_{\odot}$ to  0.0095  at a
luminosity  of 3.3$\times$10$^{12}$  L$_{\odot}$ and  the  11.3 $\mu$m
feature varies  from 0.012  to 0.004 over  the same  luminosity range.
These values  agree well with  the observational ones.  To  derive the
conversion factor  for each object,  we adopt the template  that gives
the  closest  aromatic flux  at  the  redshift  of this  object.   The
uncertainties are assumed  to be the observed ones  (40\% and 30\% for
$L$(7.7$\mu$mPAH)/$L$(8-1000$\mu$m)                                 and
$L$(11.3$\mu$mPAH)/$L$(8-1000$\mu$m), respectively), although there is
only a 10\% difference between conversion factors for SED templates in
two  adjacent  luminosity  ranges.    The  final  uncertainty  of  the
aromatic-derived  SFIR  luminosity  includes  that of  the  conversion
factor and the measurement uncertainty  of the aromatic flux.  If this
final  uncertainty is  larger  than the  measured  aromatic flux,  the
3$\sigma$  upper limit is  adopted.  Table~\ref{Quasar_PAH}  lists the
SFIR luminosity  calculated in the  above way.  For objects  with both
features detected, we  adopted the value from the  11.3 $\mu$m feature
since  it  generally has  smaller  uncertainty.   The  value from  the
detected  feature is  listed if  only  one feature  is detected.   For
objects with  neither feature  detected, the lower  value for  the two
upper limits is listed.

 As  discussed in \S~2.2,  PG 2304+042  and 3C  272.1 have  thermal IR
emission outside the IRS slit.  This extended emission is converted to
the total  IR luminosity by multiplying  by a factor of  12.0 based on
the   star-forming   template  with   $L_{IR}$(8-1000$\mu$m)=10$^{11}$
L$_{\odot}$ from  \citet{Dale02}, and is  close to the  observed value
\citep{Chary01}.

Non-star-formation  sources,  such  as planetary  nebulae  and
diffuse  stellar  radiation,  can  excite low-level  IR  emission  and
aromatic features.  Aromatic  features have  been observed  in a
fraction of elliptical galaxies \citep{Bressan06} and some of them may
originate from star  formation regions while others may  be excited by
an  old  stellar  population.   In  five  normal  elliptical  galaxies
observed by  \citet{Kaneda05}, the 11.3 $\mu$m  aromatic luminosity is
between 10$^{5}$ and 8$\times$10$^{6}$ L$_{\odot}$ \citep[the possible
problem in this work with  stellar light subtraction should not affect
the  11.3 $\mu$m  flux  much;][]{Bregman06}. To be sure we are measuring
recent star formation,  we  adopt  a  limiting
aromatic luminosity  of 3$\times$10$^{7}$ L$_{\odot}$  above which the
old stellar population contribution  should be smaller than 25\%.  The
corresponding aromatic-derived  total IR  luminosity at this  limit is
3$\times10^{9}$ L$_{\odot}$.  Therefore, a total of twenty-two objects
including eight PAH-detected ones are excluded.

\section{Origin of the Far-IR emission of AGN}

\begin{figure}
\epsscale{1.2}
\plotone{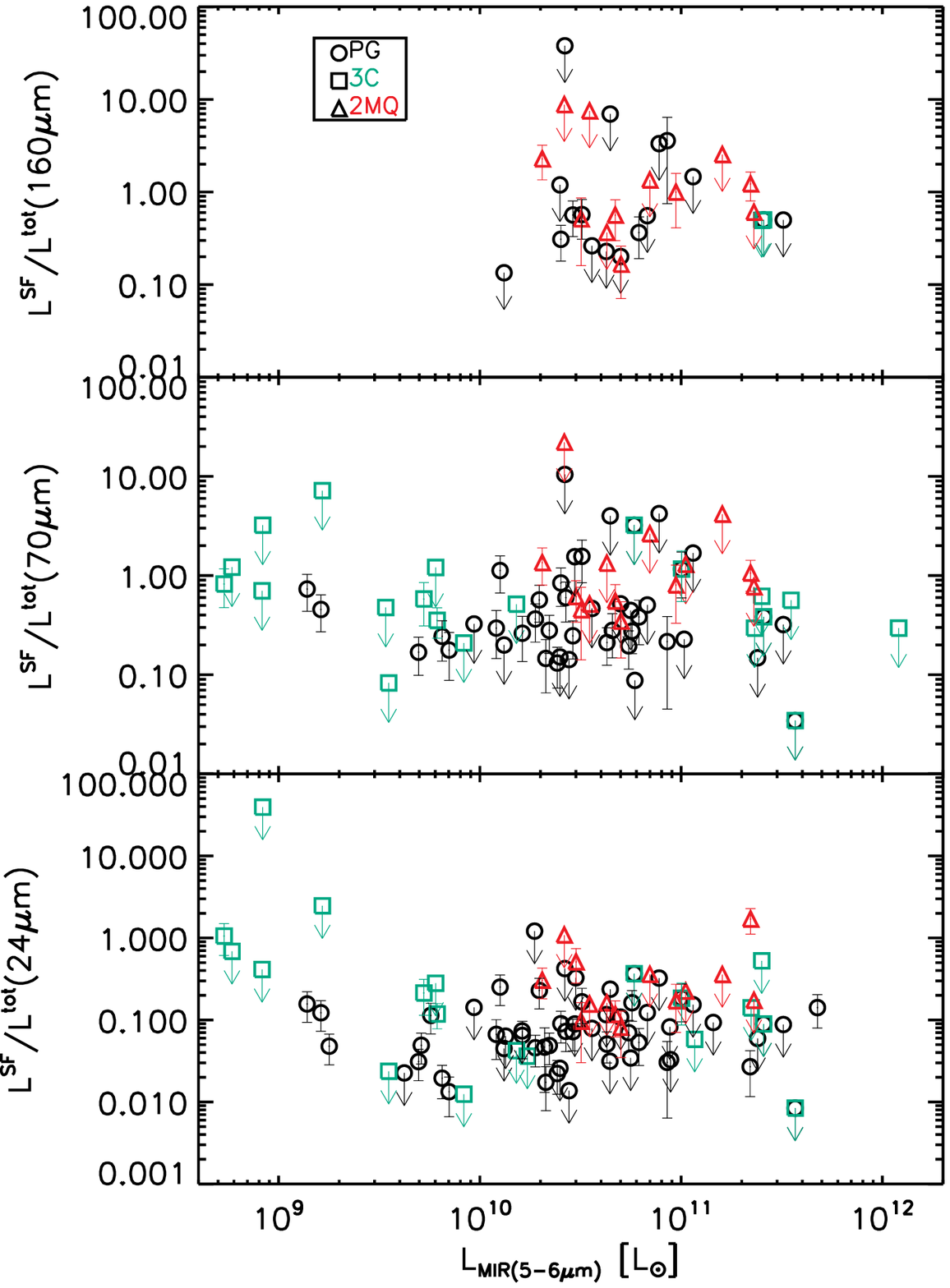}
\caption{ \label{PAH_FIR} 
The star-formation fraction at   24,  70 and 160  $\mu$m  versus the 
mid-IR (5-6$\mu$m) luminosity for the PG, 3C and 2MASS objects, respectively 
(see the color version online). }
\end{figure}

\begin{figure}
\epsscale{1.2}
\plotone{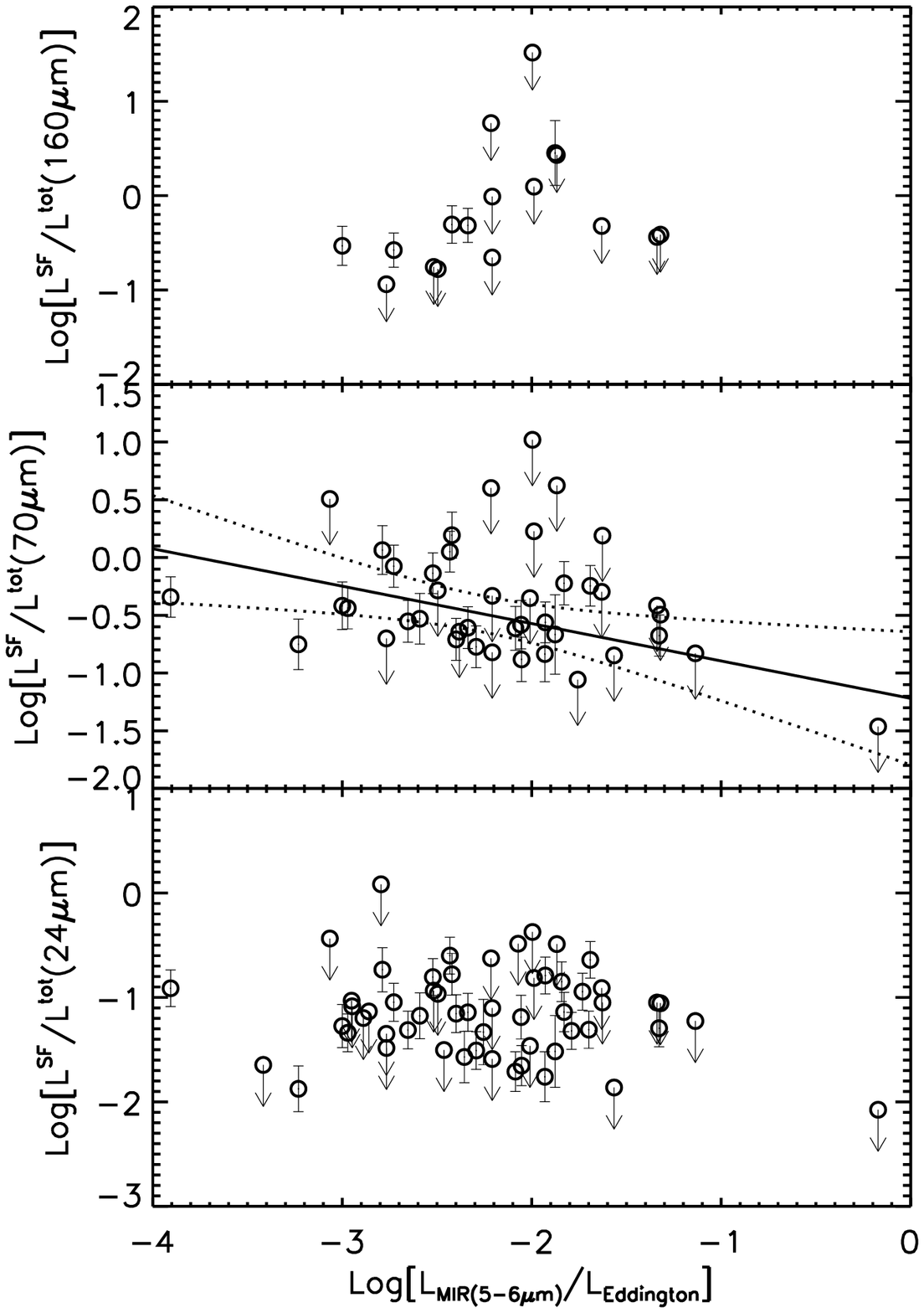}
\caption{ \label{PAH_FIR_Edd} The star-formation fraction at 
  24,  70 and 160  $\mu$m versus the ratio
of mid-IR (5-6 $\mu$m) luminosity and the Eddington luminosity  for PG quasars. The solid line
is the regression line and the two dotted lines are 2$\sigma$ confidence bounds. }
\end{figure}

Fig.~\ref{PAH_FIR} shows  the star-formation contribution  to the MIPS
rest-frame 24, 70 and 160 $\mu$m emission versus the integrated mid-IR
luminosity between 5.0 and 6.0 $\mu$m.  The {\it IRAS} or {\it ISO} 25
$\mu$m  fluxes are  plotted for  objects without  MIPS 24  $\mu$m flux
measurements.  For  objects without MIPS 70  $\mu$m flux measurements,
we estimate  one by interpolating  between the detected {\it  IRAS} or
{\it ISO} 60  and 100 $\mu$m fluxes.  The  MIPS fluxes are K-corrected
by  assuming  $\alpha$=1  for  24  and  70  $\mu$m  photometry,  and
$\alpha$=0.0   for   160   $\mu$m  photometry   (f$_{\nu}$   $\propto$
$\nu^{-\alpha}$), based  on the  IR SED of  AGN in  \citet{Haas03} and
\citet{Shi05}.  The total  PAH-derived SFIR luminosities are converted
to  the  star-formation  emission   at  the three  MIPS  bands  using  the
luminosity-dependent conversion factors  derived from the star-forming
templates from \citet{Dale01} and \citet{Dale02}.

At  24 $\mu$m, Fig.~\ref{PAH_FIR}  indicates most  of the  objects are
dominated by AGN emission.  At  70 and 160 $\mu$m, the far-IR emission
of an  individual AGN  can be  dominated by either  AGN power  or star
formation. To  quantify the star-formation fraction at  the three MIPS
bands  and its  possible dependence  on  the AGN  luminosity, we  have
employed  the code  written by  \citet{Kelly07} that  incorporates the
upperlimit  measurements.   As   listed  in  Table~\ref{SF_MIPS},  the
average star-formation fractions  for the whole sample at  MIPS 24, 70
and 160  $\mu$m are  4\%, 26\% and  28\%, respectively, at  the median
mid-IR luminosity (2.6$\times$10$^{10}$ L$_{\odot}$) of the sample. As
indicated by  Table~\ref{SF_MIPS}, these ratios  depend on luminosity,
with a lower relative star-formation contribution at higher AGN mid-IR
luminosity. The  diverse nature of far-IR emission  is consistent with
the large scatter  of the correlation between the  far-IR emission and
AGN      power     indicators      \citep[e.g.][]{Shi05,     Cleary06,
Tadhunter07}.  There will also  be some  scatter due  to the  range of
redshifts. However,  since the redshifts  of our PG and  2MASS samples
are similar and modest, the effect should be small.

Table~\ref{SF_MIPS} also includes the  result for PG and 2MASS objects
at the  MIPS 24 and 70  $\mu$m bands, where there  are enough detected
data  points.  The  average  star-formation contributions  at MIPS  70
$\mu$m  for  PG  and  2MASS   are  24\%  and  51\%  at  median  mid-IR
luminosities      of     3.0$\times$10$^{10}$      L$_{\odot}$     and
3.5$\times$10$^{10}$ L$_{\odot}$, respectively.   The fraction for the
PG   quasars    is   lower    than   that   ($>$30\%)    obtained   by
\citet{Schweitzer06}, who also employ the aromatic feature to evaluate
the role of star  formation.  Contributions to the discrepancy include
a difference in the conversion factors from the aromatic fluxes to the
SFIR fluxes and the relatively large uncertainties in their 7.7 $\mu$m
fluxes caused by silicate features, whereas our result is mainly based
on 11.3 $\mu$m features.

Compared  to the  whole sample,  PG objects  show  relatively stronger
luminosity-dependence  of the  star-formation fractions  at 24  and 70
$\mu$m,  with  decreasing  fractions  at higher  mid-IR  luminosities.
However, the 2MASS objects do not have such a relation and most of the
3CR results are  upperlimits.  Thus the relation for  the whole sample
is   mainly    produced   by   the    PG   sample.    As    shown   in
Fig.~\ref{PAH_FIR_Edd} and Table~\ref{SF_MIPS_Edd}, the star-formation
fractions for the PG objects also decrease as the ratios of the mid-IR
continuum luminosities and  the Eddington luminosities decrease, where
the   blackhole    masses   of   PG   objects    are   obtained   from
\citet{Vestergaard06}  and   \citet{Kaspi00}.   The  anti-correlations
indicate these  two relations are  not caused by the  selection effect
that the  detectable aromatic features  in objects with  higher mid-IR
continuum emissions have larger fluxes.

\section{STAR-FORMING IR LUMINOSITY FUNCTION OF QUASAR HOST GALAXIES}

\subsection{Methodology}

The main challenge in deducing the SFIR luminosity function (LF) for our
sample is  that the  flux limit  of the aromatic  feature is  not well
defined and many objects have only upper limits in these measurements.  Therefore, we
obtained  the SFIR  LF  by  converting the  well-defined  LF at  other
wavelengths      using      the      fractional      bivariate      LF
\citep{Elvis78}. The formula can be written as
\begin{equation} 
\Phi_{M_{\rm SFIR}} = \sum_{M_{\lambda}}\Phi_{M_{\lambda}}F(M_{\lambda}, M_{\rm SFIR})
\end{equation}
where $\Phi_{M_{\rm SFIR}}$ is the SFIR LF and $\Phi_{M_{\lambda}}$ is
the LF at  $\lambda$-band where each parent sample  is selected (radio
for  3CR objects,  $B$-band  for  PG objects  and  $K$-band for  2MASS
objects). The  fractional bivariate LF  $F(M_{\lambda}, M_{\rm SFIR})$
indicates  the fraction  of  objects with  magnitude $M_{\lambda}$  at
$\lambda$-band having SFIR luminosity of $M_{\rm SFIR}$.  We calculate
$F(M_{\lambda},   M_{\rm   SFIR})$${\Delta}M_{\lambda}$${\Delta}M_{\rm
SFIR}$ by  dividing the number $n_{1}$ of  objects with $\lambda$-band
magnitude in  the interval $M_{\lambda}{\pm}{\Delta}M_{\lambda}/2$ and
the SFIR  luminosity in the  interval $M_{\rm SFIR}{\pm}{\Delta}M_{\rm
SFIR}/2$  by  the  number   $n_{2}$  of  objects  with  $\lambda$-band
magnitude in the interval $M_{\lambda}{\pm}{\Delta}M_{\lambda}/2$ that
could  have   had  detected  aromatic   features  if  they   had  SFIR
luminosities of $M_{\rm SFIR}$.   $n_{1}$ is the observed number.  Any   
object   with   $\lambda$-band   magnitude   in   the   interval
$M_{\lambda}{\pm}{\Delta}M_{\lambda}/2$ will  be counted into $n_{2}$,
if it  has a limiting SFIR  luminosity lower than  $M_{\rm SFIR}$. The
limiting SFIR luminosity is defined as the minimum star formation rate
to  detect the  aromatic feature  (see  \S~2.3) plus  any extended  IR
emission.

For PG quasars, $\Phi_{M_{\lambda}}$ is the $B$-band LF at 0.0$<z<$0.5
from  Table 9  of \citet{Schmidt83},  where  the median redshift  of 0.25  is
adopted to  convert the apparent  magnitude to the  absolute magnitude
and the  K-correction is the  same as described  in \citet{Schmidt83}.
This $B$-band LF has data coverage for $M_{B}$ from -21.4 mag to -26.4
mag.    A    double-exponential   model   \citep[for    the   formula,
see][]{LeFloch05} fits the  $B$-band LF well and it  is used to derive
the $\Phi_{M_{B}}$  for any given  $M_{B}$ between -21.0 and  -26.5 for
our PG subsample.  The SFIR  luminosity of this PG subsample spans the
range  from  3.1$\times10^{9}$  to 2.4$\times10^{12}$  L$_{\odot}$.   To
construct the fractional  bivariate LF ($F$($M_{B}$, $M_{\rm SFIR}$)),
the entire ranges of $M_{B}$ and SFIR luminosity are each divided into
four  intervals.   The  final  fractional bivariate  LF  ($F$($M_{B}$,
$M_{\rm  SFIR}$)) along  with  Poissonian uncertainties  is listed  in
Table~\ref{FB_PG}.

For 2MASS objects, the LF  at $K$ band from \citet{Cutri01} is adopted
as  $\Phi_{M_{\lambda}}$.  A  two-exponential model  does not  fit the
data well and thus we interpolate  the measured data points to get the
space density  at a given $K$-band  magnitude. Table~\ref{FB_2M} lists
the final  fractional bivariate LF ($F$($M_{B}$,  $M_{\rm SFIR}$)) for
2MASS objects.

For  3CR objects,  $\Phi_{M_{\lambda}}$  is  the LF  at  151 MHz  from
\citet{Willott01}, where the LF is obtained based on the 3CRR, 6CE and
7CRS samples.   We use the analytic  LF of model C  for a cosmological
model   of   $\Omega_{m}$=0,   $\Omega_{\lambda}$=0   and   $H_{0}$=50
kms$^{-1}$Mpc$^{-1}$, because  the LF  for this cosmological  model is
close to that for our cosmological model except for the $H_{0}$ value.
\citep{Willott01}.  We  convert to  our cosmological model  by setting
$\Phi_{1}(L_{1},     z)dV_{1}      =     \Phi_{2}(L_{2},     z)dV_{2}$
\citep{Peacock85}.   The  radio luminosity  at  151  MHz  for our  3CR
subsample  is calculated and  K-corrected using  the flux  density and
spectral index at 178 MHz  from \citet{Spinrad85}. Again, we limit our
3CR subsample to  the redshift range between 0.0 and  0.5 to match the
PG  and 2MASS  redshift  ranges.  The  final  fractional bivariate  LF
($F$($M_{151MHz}$, $M_{\rm SFIR}$))  with Poissonian uncertainties is
listed in Table~\ref{FB_3C}.

\subsection{Star-forming IR Luminosity Function of Active Galaxies}
\subsubsection{Comparison to Field Galaxies}

\begin{figure}
\epsscale{1.3}
\plotone{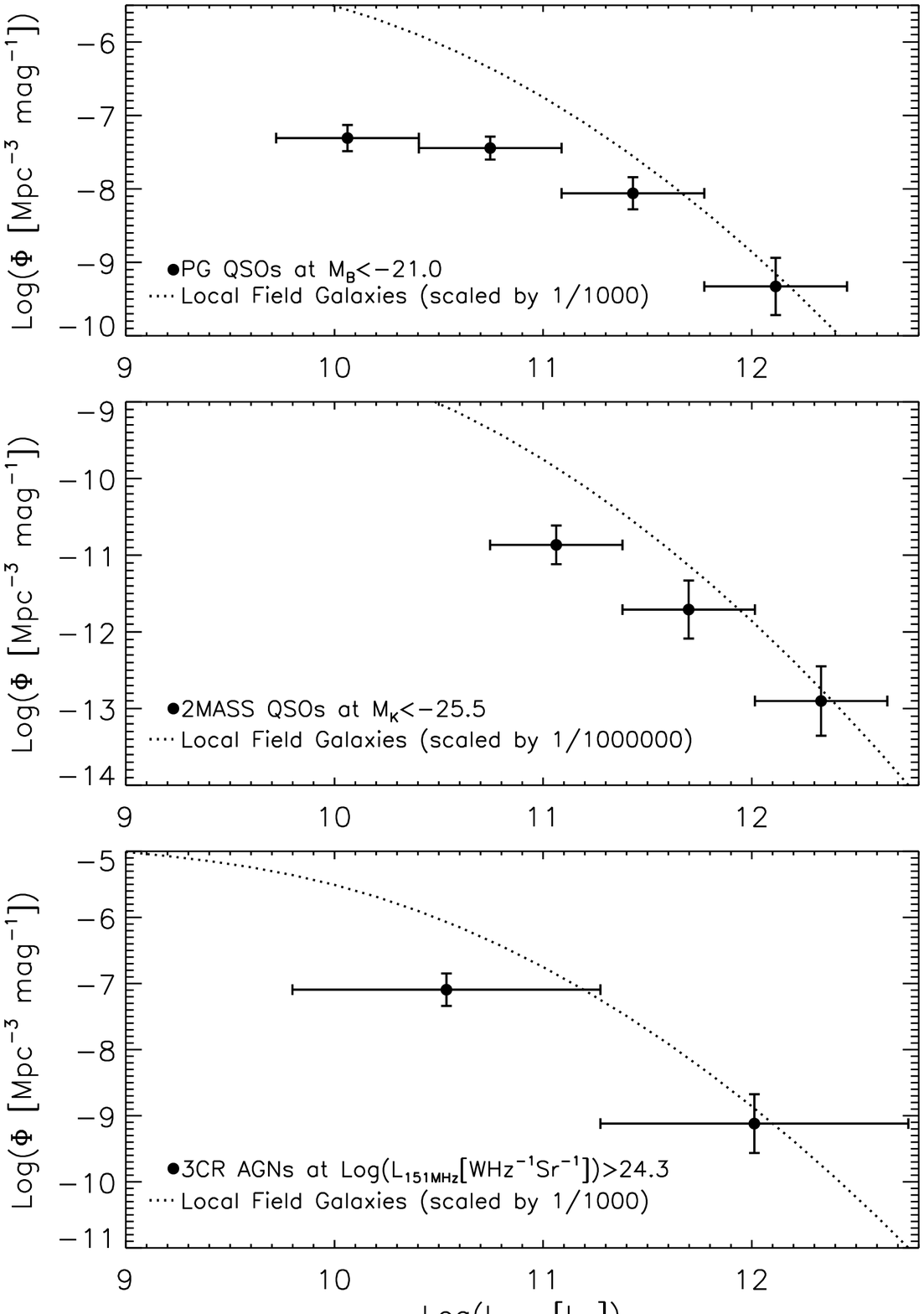}
\caption{
\label{LF_totIR_PAH} Star-forming infrared luminosity functions for the PG, 2MASS and
3CR AGN.   The dotted  line is the  re-normalized luminosity  function of
local field  galaxies from \citet{LeFloch05}.   }
\end{figure}

The  most  important  result  from  the fractional  bivariate  LFs  in
Table~\ref{FB_PG},  Table~\ref{FB_2M}  and  Table~\ref{FB_3C} is  that
objects with  a large range of  nuclear activity have  a non-zero
probability  of  having  a high   SFIR  luminosity.   The  form  of  the
fractional bivariate LF  implies that SFIR LF of  AGN host galaxies is
much flatter than the LF of the AGN themselves.

Fig.~\ref{LF_totIR_PAH} shows the results for  the SFIR LF for the PG,
2MASS and 3CR subsamples. Each subsample has a brightness limit at the
wavelength  where  it is  selected.  We  set  $M_{B}<$-21 for  the  PG
subsample  and  $M_{K}<$-25.5  for  the 2MASS  subsample  and  $L_{\rm
151MHz}>$2$\times$10$^{24}$   W  Hz$^{-1}$   Sr$^{-1}$  for   the  3CR
subsample.  The  dotted line  shows the re-normalized  IR LF  of local
field galaxies from \citet{LeFloch05} based on the {\it IRAS} and {\it
ISO} results;  it agrees well  with previous studies  of the IR  LF of
field     galaxies      \citep[See][]{Rieke86,     Sanders03}.      In
Fig.~\ref{LF_totIR_PAH},  the SFIR  LFs of the three subsamples  are much
flatter than the re-normalized LFs of field galaxies.

We need to be  sure that the flatter LFs are not  just a result of the
difficulty in measuring the SFR  around a bright quasar.  We first use
Monte-Carlo simulations to test the robustness of the methodology used
to  derive the  SFIR LF  of AGNs.   The following  steps are  taken to
construct  a sample  that  mimics the  PG  subsample: (1)  a total  of
$N_{obj}$($>$10000) objects is created over the redshift range between
0.001 and  0.5; (2) the comoving  number density is  constant over the
redshift range; (3)  a $B$-band luminosity within the  range of the PG
subsample  is  assigned  to  each  object randomly  but  the  relative
distribution is  the same as the  PG $B$-band LF;  (4) similarly, each
object   has  a   randomly  assigned   IR  luminosity   with  relative
distribution  defined by the  SFIR LF  of the  PG subsample.   In this
case, the  IR flux  is not  correlated with the  $B$-band flux;  (5) a
well-defined flux limit  is applied in the $B$-band  while the IR flux
limit is randomly distributed over  the whole range of the SFIR fluxes
of the  PG subsample.  After producing  the above set  of objects, the
fractional bivariate LF is  calculated based on those objects detected
in  the $B$-band.   The final  derived  SFIR LF  using the  fractional
bivariate  LF follows  the  pre-defined SFIR  LF  within the  Poission
noise.  The same result is obtained for the simulation in which the IR
flux is tightly correlated with the $B$-band flux.

Unlike  the  PG  subsample,  which  is  complete,  the  2MASS  and  3CR
subsamples only contain one-third  of their parent samples at $z<$0.5.
To test for the effects of  the sample incompleteness, we use only one
third of the objects brighter  than the $B$-band limiting flux created
in  the above  simulations, with  these objects  having  the brightest
apparent $B$-band magnitude.  Again, the derived SFIR LF is consistent
with pre-defined SFIR  LF within the Poission noise.   We also test
using the one-third  of the objects  with the most luminous  absolute $B$-band
luminosity.  The  shape of the derived  IR LF does not  change but the
normalization  becomes smaller.  The  same result  is obtained  if the
$B$-band  flux  correlates with  the  IR  flux.  Thus, for  all  three
subsamples, the  Monte Carlo code  demonstrates the robustness  of our
methodology to derive the SFIR LF of AGNs

Because  AGNs  have  strong  mid-IR continua,  aromatic  features  are
detected only  in host galaxies  with intense star formation.   We can
now use the Monte-Carlo  simulation to demonstrate that this selection
effect cannot account for the  large difference in the SFIR LF between
the field  galaxy and  PG quasars.  In  the simulation, we  assume the
SFIR LF  of PG quasars actually  follows that of  field galaxies.  For
each PG object, we obtain the  IR LF of field galaxies at the redshift
of this  object by  assuming that  the local field  galaxy IR  LF from
\citet{LeFloch05}    evolves   with    redshift    as   $L^{*}(z)    =
L^{*}(0)(1+z)^{3.2}$ and $\Phi^{*}(z)=\Phi^{*}(0)(1+z)^{0.7}$. We then
randomly assign  a SFIR luminosity to  this PG object  with a relative
probability that  follows the LF  of field galaxies at  this redshift.
The   range    of   the   simulated   SFIR    luminosities   is   from
3.1$\times$10$^{9}$  to 2.4$\times$10$^{12}$  L$_{\odot}$, consistent
with the observed range for the  PG quasars.  Also, we assume that the
total probability  in this  luminosity range is  equal to 1.   In this
case, all simulated  IR luminosities are above the  low luminosity cut
(3$\times$10$^{9}$ L$_{\odot}$),  and thus bias  the results toward
the high luminosity end.   Combining the simulated SFIR luminosity and
the observed  uncertainty or  upper limit for  each PG object,  we can
calculate the detection fraction for the aromatic features.  After one
thousand  simulations,   we  find   (despite  the  bias   toward  high
luminosity)  that the  detection fraction  is only  (28$\pm$3)\%, much
smaller  than  the  observed  value  (48\%).   This  large  difference
indicates that  our result is not  simply due to  the selection toward
high levels of SFR caused by the AGN emission.

We further measure the probability of producing the observed curvature of the
SFIR LF if the PG quasar sample actually  has a field galaxy SFIR LF. In each
simulation, all PG objects  are assigned randomly SFIR luminosities as
described above.   Using the  simulated luminosities and  the observed
uncertainties or upperlimits, a SFIR  LF is constructed using the same
procedure  including the  number of  luminosity bins  as  the observed
LF. All  data points produced in  a total of  ten thousand simulations
are  rebinned to the  same bins  as for  the observed  PG LF.  In four
luminosity bins,  the fractions of simulated  non-zero number densities
are  100\%, 100\%, 64\%  and 6\%  from low  to high  luminosity.  All
simulated number densities are then  rescaled by a factor to match the
composite number density  in the first luminosity bin  to the observed
one.  This composite number density  is assumed to be the median value
of all simulated number densities  (including zero value) in the first
bin,  indicating  a  probability  of  50\%.   We  then  calculate  the
probability  for  an  observed  luminosity  bin  as  the  fraction  of
simulated number densities larger than  the lower 1-sigma bound of the
observed number density in this bin.  The probability in each bin from
low   to  high  luminosity   is  99.0\%,   1.0\%,  2.5\%   and  4.0\%,
respectively.  This result provides  further evidence that the flatter
SFLF of the PG quasars is robust against  selection effects.

\subsubsection{Dependence  on AGN Luminosity}

\begin{figure}
\epsscale{1.2}
\plotone{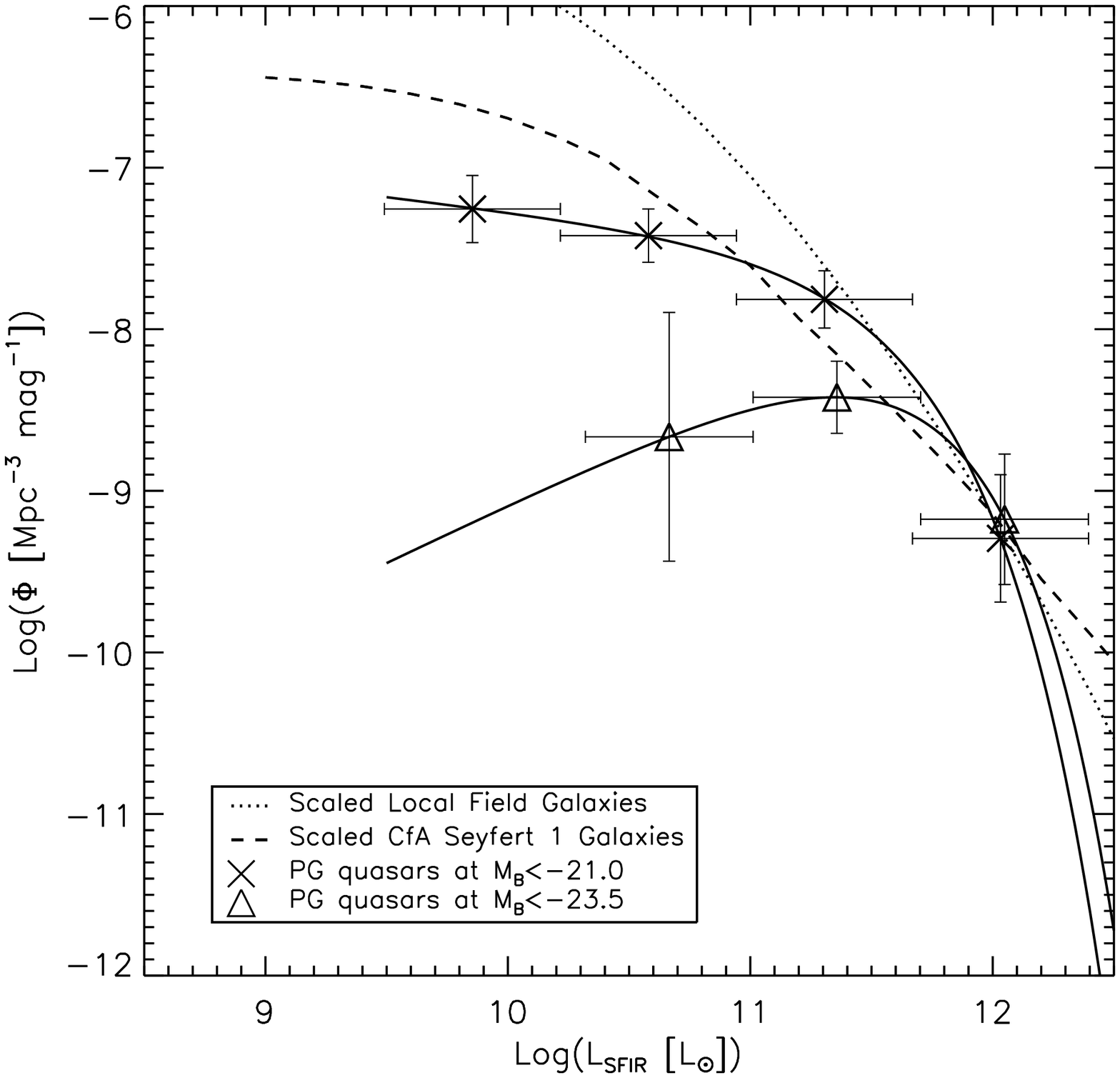}
\caption{ \label{LF_totIR_PAH_QSOBRIGHT} 
Star-forming infrared luminosity functions of PG quasars as a function
of quasar brightness. The  dashed line is the re-normalized luminosity
function   of  star  formation   in  CfA   Seyfert  1   galaxies  from
\citet{Maiolino95}.  The dotted line  is the  re-normalized luminosity
function  of local  field galaxies  from \citet{LeFloch05}.  The solid
lines are Schechter-function fits to the two PG subsamples. }
\end{figure}

Fig.~\ref{LF_totIR_PAH_QSOBRIGHT} shows the SFIR LF of PG quasars as a
function  of  the  $B$-band  luminosity.   The  two  solid  lines  are
Schechter-function fits for PG quasars at $M_{B}<$-21 and $M_{B}<$-23,
respectively.     The     fitting    parameters    are     given    in
Table~\ref{Best_Fit}.  There is a trend that the SFIR LF of PG quasars
becomes  flatter for  the brighter  PG objects.   We suggest  that the
higher SFR for  brighter PG quasars is not  a selection effect because
the  $B$-band  luminosity of  normal  infrared  galaxies  is not  well
correlated with  IR luminosity and  LIRGs rarely have M$_{B}$  $<$ -23
\citep[See][]{Rieke86}.         The        trend        seen        in
Fig.~\ref{LF_totIR_PAH_QSOBRIGHT} is not likely to be due to evolution
with  redshift,  as  the  mean  redshifts for  the  faint  and  bright
subsamples are nearly the same from faint to bright, 0.18$\pm$0.30 and
0.24$\pm$0.11 respectively.

In  Fig.~\ref{LF_totIR_PAH_QSOBRIGHT}, the  dashed line  is the  LF of
extended   star   formation   in   CfA   Seyfert   1   galaxies   from
\citet{Maiolino95}.  The extended IR  emission of Seyfert galaxies was
obtained  by subtracting  the  nuclear emission  from  IRAS 12  $\mu$m
photometry  \citep[See][]{Maiolino95}.   We  converted the  10  $\mu$m
luminosity to the  total IR luminosity using the  IR SED template from
\citet{Dale01}  and  \citet{Dale02}.    Similarly  to  converting  the
aromatic flux to  the total IR luminosity, the  conversion factor from
10  $\mu$m  to  the  total  IR  luminosity depends  on  the  total  IR
luminosity.  The omission of  nuclear star formation (within 2$''$) in
the  study of  \citet{Maiolino95}  may  affect the  LF  of total  star
formation in their Seyfert galaxies.   However, if nuclear star formation is
correlated   with   the   extended   star  formation   as   found   by
\citet{Buchanan06}, the shape  of the LF for the  total star formation
in  Seyfert    galaxies   should    not   change.    As    shown   in
Fig.~\ref{LF_totIR_PAH_QSOBRIGHT}, the  SFIR LF of  Seyfert 1 galaxies
is steeper than the LF of PG quasars.  There is also a suggestion that
the LF  for the  lower-luminosity PG quasars  is steeper than  for the
higher-luminosity  ones.   Seyfert  galaxies  have a  higher  SFR  and
flatter    LF   on    average   than    field    galaxies   \citep[see
Fig.~\ref{LF_totIR_PAH_QSOBRIGHT} and][]{Maiolino95}.  It appears that
star formation is  correlated with the level of  nuclear activity over
the full range from normal galaxies to quasars.

To test the trend  of the SFIR LF of active galaxies  as a function of
AGN luminosity,  we extended the Monte-Carlo  simulations described in
\S~6.2.1 to  test the difference  between PG quasars  with $M_{B}<$-21
and PG quasars  with $M_{B}<$-23.  In this simulation,  we assume that
the SFIR LF of PG quasars with $M_{B}<$-23 actually follows that of PG
quasars with $M_{B}<$-21. For a  PG quasar with $M_{B}<$-23, we obtain
the SFIR  LF of PG  quasars with $M_{B}<$-21  at the redshift  of this
object by assuming  the SFIR LF of PG  quasars at $M_{B}<$-21 evolving
with redshift  as $L^{*}(z) = L^{*}(z_{1})(\frac{1+z}{1+z_{1}})^{3.2}$
and   $\Phi^{*}(z)=\Phi^{*}(z_{1})(\frac{1+z}{1+z_{1}})^{0.7}$,  where
$z_{1}$ is  the mean  redshift (0.2) of  PG quasars  with $M_{B}<$-21.
Based on this LF, a random  SFIR luminosity is assigned to a PG quasar
with $M_{B}<$-23.  The luminosity range is between 3.1$\times$10$^{9}$
and 2.4$\times$10$^{12}$,  consistent with  the observed range  for PG
quasars with  $M_{B}<$-21.  The  total probability in  this luminosity
range  is equal  to  1.   Using the  observed  uncertainties or  upper
limits, we predict the detection  fraction of the aromatic feature for
PG  quasars at $M_{B}<$-23  of 17$\pm$5\%,  smaller than  the observed
fraction of  28\%. This result  supports our conclusion that  the SFIR
luminosity increases with increasing AGN luminosity.

\subsubsection{Comparison Between Different Subsamples}

As shown in Fig.~\ref{LF_totIR_PAH}, the behavior of star formation is
different around AGN selected by different techniques.  Since the SFIR
LF of  AGN host galaxies is a  function of AGN luminosity  as found in
the last  section, the  effect of the  nuclear brightness needs  to be
removed.   The  2MASS  $K$-band  photometry  for all  PG  objects  was
obtained from the 2MASS Point Source Catalog.  We calculated $B-K$ for
all  PG  objects and  found  that  $<B-K>$=3.0$\pm$0.6  and is  not  a
function  of  absolute  $K$-band   magnitude.   All  PG  objects  with
$M_{B}<$-22.5 are selected  to form a comparison sample  for the 2MASS
objects with $M_{K}<$-25.5.  For the 3CR subsample, it is difficult to
select a PG  sample with the same level of  nuclear activity.  This is
because PG objects are selected  by thermal emission while 3CR objects
are selected  because of  their non-thermal emission  and there  is no
good  correlation between the  radio emission  and the  thermal mid-IR
emission \citep{Ogle06}.   Instead, we compare the  whole PG subsample
at     $M_{B}<$-21     to    the     whole     3CR    subsample     at
$L_{151MHz}>$2$\times$10$^{24}$ W Hz$^{-1}$ Sr$^{-1}$.

\begin{figure}
\epsscale{1.2}
\plotone{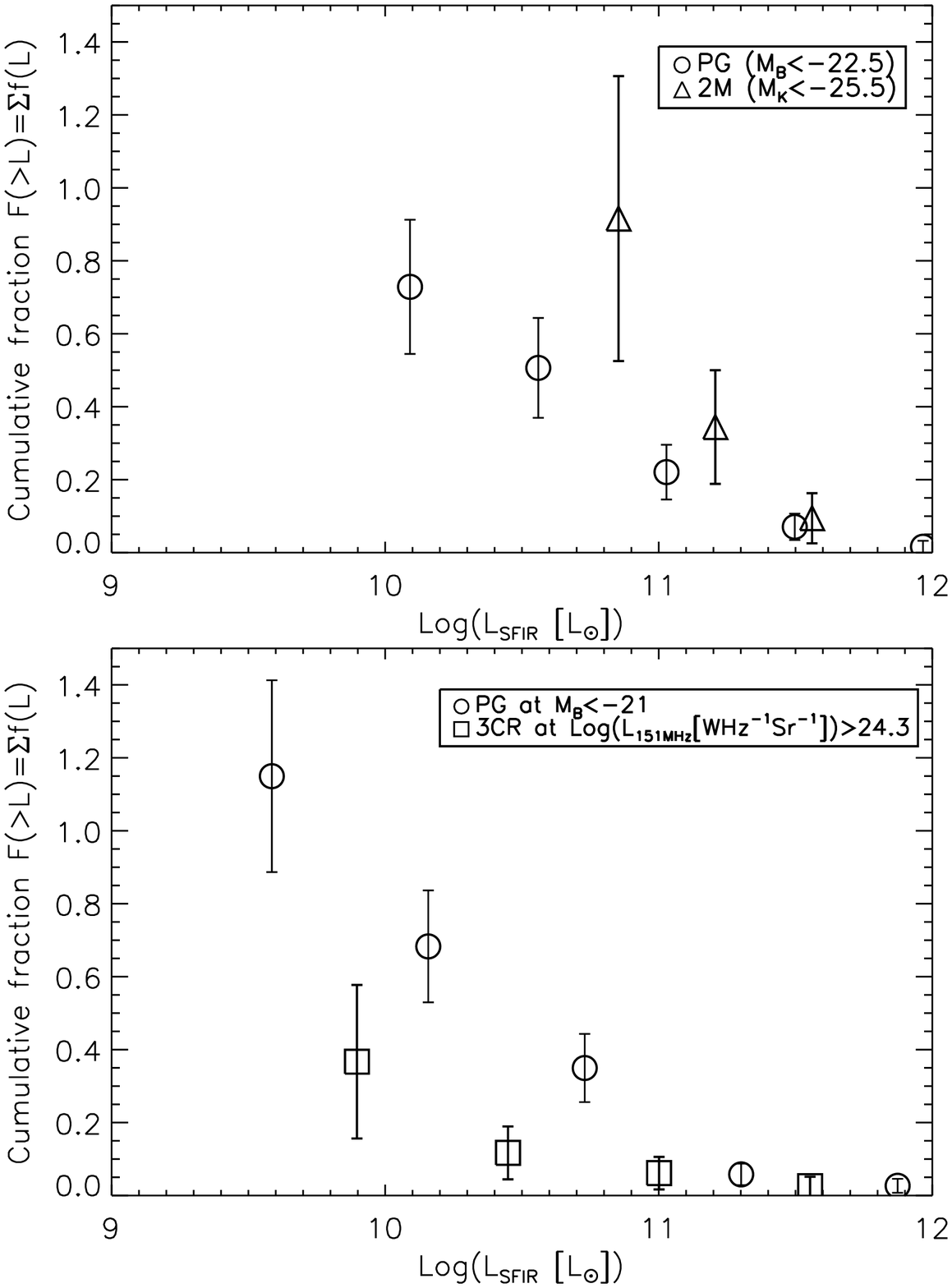}
\caption{\label{Frac_FL_PAH}  
Cumulative     fraction     luminosity     functions     F($>$L)     =
$\sum_{L=L_{0}}^{\infty}$ f(L)  for the PG objects versus  2MASS  objects (upper plot) and the PG objects versus 3CR
objects (lower plot), where   f(L)  is  the  fractional
luminosity function (See text).}
\end{figure}

Fig.~\ref{Frac_FL_PAH}  shows  the  cumulative  fractional  luminosity
function F($>$L) = $\sum_{L=L_{0}}^{\infty}$  f(L) for PG versus 2MASS
and PG versus 3CR. To avoid biases due to evolution, the comparison includes objects
with $z<$0.5.  The fractional luminosity function f(L) is defined
similarly  to  the   fractional  bivariate  LF  \citep[See][]{Elvis78,
Golombek88}.  As shown in Fig.~\ref{Frac_FL_PAH}, there is an apparent
sequence in terms  of the level of SFR that progresses  from 3CR to PG
to 2MASS  objects that  generally show the  highest SFRs.   The median
star-forming  IR   luminosities  of  3CR,  PG  and   2MASS  objects  are
6$\times$10$^{9}$,     3.0$\times10^{10}$    and    1$\times$10$^{11}$
L$_{\odot}$,  respectively.   Different  AGN  selection  techniques
appear  to identify  objects  with different  levels  of star  forming
activity in their host galaxies.

\subsection{Implications for Nuclear Activity}

The  flatter   SFIR  LF  of  AGN  host   galaxies  indicates  enhanced
star-forming  activity  relative to  local  field galaxies.   Previous
studies illustrate the  presence of significant post-starburst stellar
populations  in quasar  host galaxies.  For example,  the  optical and
near-IR broadband SEDs  of AGN indicate the presence  of young stellar
populations  with  an  age  of  about  a Gyr  in  the  host  galaxies,
independent  of morphological  type \citep{Jahnke04},  consistent with
previous  studies   \citep{Kotilainen94,  Schade00,  Ronnback96}.   In
addition, \citet{Kauffmann03}  found a  trend of younger  mean stellar
population  for higher-luminosity AGN  based on  a very  large sample.
None of these studies found evidence for intense on-going massive star
formation,  except  for  a  few objects  \citep[see][]{Jahnke04}.   We
emphasize that the techniques employed in the above studies are unable
to  detect OB  stars  or  suffer from  strong  degeneracy between  the
current  star-formation and  the  star-formation history.   Therefore,
these studies do not contradict our result.  Searches for massive star
formation through  UV spectroscopy or  spatially-resolved observations
for  star-formation  tracers  (such  as  recombination  lines  and  IR
emission) indicate  the presence of massive star  formation in Seyfert
galaxies     \citep{Maiolino95,    Heckman97}    and     in    quasars
\citep{Cresci04}.  All of these studies focus on the central region of
the  galaxy,   implying  that  the   star  formation  in   quasars  is
circumnuclear.   This is  consistent  with the  lack of  spectroscopic
evidence for on-going  star formation at distances from  the nuclei of
$\sim$15 kpc \citep{Nolan01}.

The flatter  SFIR LF of AGN  host galaxies relative  to field galaxies
also implies that  nuclear activity tends to be  triggered in galaxies
with enhanced star formation. Based on Fig.~\ref{LF_totIR_PAH}, we can
calculate the probability of triggering  a PG quasar in field galaxies
at a  given SFR;  for example, the  probability of  triggering nuclear
activity at  $L_{SFIR}$=1.25$\times$10$^{12}$ L$_{\odot}$ is  a factor
of 50  higher than that  at $L_{SFIR}$=1$\times$10$^{10}$ L$_{\odot}$.
This  indicates  an environment  with  intense  star formation  offers
preferential  conditions for  nuclear activity,  such as  the abundant
inflowing material driven by star formation \citep{Granato04}.  On the
other  hand, it implies  that over  much of  the life  of an  AGN, its
feedback does not  quench the star formation, but  instead may enhance
the  host galaxy  star  formation as  demonstrated  in some  numerical
simulations \citep{Silk05}.   Our result  that more luminous  AGNs are
more  likely  to  reside  in  host galaxies  with  more  intense  star
formation  provides  further  evidence  that  feedback  from  the  two
physical processes  (star formation and nuclear  activity) can enhance
both processes.  Numerical simulations have predicted the evolution of
the   SFR  and  SMBH   accretion  rate   along  the   merging  process
\citep{Granato04,  Springel05}.  They conclude  that the  evolution of
star formation  almost follows the  SMBH accretion rate,  although the
former  starts to  decline a  little  earlier. A  more quantative  and
careful comparison  between the simulations and  our observations will
improve our  understanding of  when and how  feedback plays a  role in
galaxy evolution and SMBH growth.

Although PG, 2MASS and 3CR AGN have flatter SFIR LFs compared to field
galaxies,  they  show differences  in  the  distribution  of SFRs,  as
indicated by the  cumulative fractional LFs in Fig.~\ref{Frac_FL_PAH}.
Fig.~\ref{PAH_CO} shows  that the SFR of AGN  host galaxies correlates
with the  amount of molecular gas  in the host  galaxy, which suggests
that  different  AGN  selection  methods  prefer  host  galaxies  with
different levels of gas reservoir. It is interesting that PG and 2MASS
quasars  have different  levels  of SFR.   Both  samples are  selected
through thermal emission.   There is no obscuration along  the line of
sight for PG  objects while the red IR-optical  color of 2MASS objects
is attributed to  the obscuration of nuclear radiation  by dust in the
circumnuclear   regions  or   host   galaxies  \citep[e.g.][]{Smith02,
Marble03}.     According     to    the    AGN     unification    model
\citep{Antonucci93},  2MASS objects  are reddened  counterparts  of PG
objects.   The different  levels of  star  formation in  2MASS and  PG
objects  suggest that  star  formation  affects our  view  of the  AGN
phenomenon, which is not  expected under the unification model.  This
is not a selection effect that 2MASS objects need to have a larger SFR
to  have comparable  the $K$-band luminosity  to PG  quasars,  as $K$-band
fluxes in 2MASS objects are dominated by hot dust or starlight, not by
star formation.   A similar correlation  has been observed  in Seyfert
galaxies, that Seyfert 2 objects have larger star formation rates than
Seyfert  1s \citep[e.g.][]{Edelson87,  Maiolino95}.   Observations and
numerical simulations show that  the feedback produced by nuclear star
formation can  heat the circumnuclear  material and thus  increase its
scale height \citep{Maiolino99, Ohsuga99, Wada02, Watabe05}. Such behavior
could produce the link between star formation activity and AGN properties.

\section{CONCLUSIONS}

We  present {\it  Spitzer} IRS  observations of  three  AGN samples
including PG quasars,  2MASS quasars and 3CR radio-loud  AGNs.  The PG
sample includes  all PG quasars at  z$<$0.5 while one  third of the
2MASS and 3CR parent samples are  used in this study. The main results
are the following:

1. The aromatic features  at 7.7 and 11.3 $\mu$m  are detected against
the strong mid-IR continuum of  the AGN.  The excitation mechanism for
the aromatic features is predominantly star formation.

2. The  contribution  of star  formation  to  the  far-IR emission  of
individual AGN is diverse; the  average contribution is around 25\% at
70  and 160  $\mu$m.   For  the PG  objects,  this contribution  shows
anti-correlations  with the  mid-IR luminosity  and the  ratio  of the
mid-IR continuum and the Eddington luminosity.

3. The star-forming  IR luminosity functions of AGNs  are flatter than
that of field galaxies, implying  the feedback from star formation and
nuclear activity can enhance both processes.

4. The star-forming IR luminosity  function of AGNs is correlated with
the  level  of nuclear  activity  over  the  whole range  from  normal
galaxies to bright quasars, with  higher star formation rates for more
intense  nuclear   activity.   The  2MASS,   PG  and  3CR   AGNs  have
distributions of star formation that follow the progression (from high
to  low  SFR)  of  2MASS-PG-3CR,  implying  that  various  AGN  survey
techniques select host galaxies  with different levels of star forming
activity.

\acknowledgements

We thank J.D. Smith for  helpful suggestions and the anonymous referee
for  detailed comments.   Support for  this work  is provided  by NASA
through  contract  1255094  and  1256424  issued  by  JPL/  California
Institute of Technology. This work  is based on observations made with
the Spitzer Space  Telescope, which is operated by  the Jet Propulsion
Laboratory, California  Institute of Technology under  a contract with
NASA.   This research  has  made use  of  the NASA/IPAC  Extragalactic
Database  (NED) which is  operated by  the Jet  Propulsion Laboratory,
California Institute  of Technology, under contract  with the National
Aeronautics and  Space Administration.  This publication  makes use of
data products  from the Two  Micron All Sky  Survey, which is  a joint
project of the University of Massachusetts and the Infrared Processing
and Analysis Center/California Institute  of Technology, funded by the
National Aeronautics and Space Administration and the National Science
Foundation.

\clearpage
\LongTables 
\begin{deluxetable}{lccccccccccccccccccccccccc} 
\tabletypesize{\scriptsize}
\tablecolumns{7}

\tablecaption{\label{Quasar_PAH} AGN with associated physical parameters }
\tablewidth{0pt}
\tablehead{
\colhead{source}               &  \colhead{Redshift}              & \colhead{F(7.7$\mu$m)}           &
\colhead{EW(7.7$\mu$m)}        &  \colhead{F(11.3$\mu$m)}         & \colhead{EW(11.3$\mu$m)}         & 
\colhead{$L_{\rm SFIR}$}       &  \colhead{$L_{5-6{\mu}m}$}       & \colhead{S$_{\rm CO}{\Delta}V$}  &  
\colhead{Ref}           \\
  \colhead{(1)}                & \colhead{(2)}               & \colhead{(3)}               & \colhead{(4)}                & 
  \colhead{(5)}                & \colhead{(6)}               & \colhead{(7)}               & \colhead{(8)}                &
  \colhead{(9)}                & \colhead{(10)}        
}
\startdata
                    PG0003+158 &0.450 &                &      &       $<$ 0.13 &      &                                    &          1.2${\times}10^{11}$ &         &   \\
                    PG0003+199 &0.025 &       $<$ 1.39 &      &  0.29$\pm$0.06 & 0.01 &    (8.8$\pm$4.03)${\times}10^{08}$ &          3.3${\times}10^{09}$ &         &   \\
                    PG0007+106 &0.089 &       $<$ 1.39 &      &  0.51$\pm$0.06 & 0.03 &    (3.2$\pm$1.32)${\times}10^{10}$ &          1.9${\times}10^{10}$ &$<$ 3.00 &  1\\
                    PG0026+129 &0.142 &       $<$ 0.36 &      &       $<$ 0.12 &      &            $<$4.5${\times}10^{10}$ &          3.0${\times}10^{10}$ &         &   \\
                    PG0043+039 &0.385 &       $<$ 0.36 &      &       $<$ 0.08 &      &            $<$5.1${\times}10^{11}$ &          1.0${\times}10^{11}$ &         &   \\
                    PG0049+171 &0.064 &       $<$ 0.50 &      &       $<$ 0.05 &      &            $<$3.1${\times}10^{09}$ &          2.1${\times}10^{09}$ &         &   \\
                    PG0050+124 &0.061 &  8.28$\pm$5.61 & 0.05 &  2.77$\pm$0.25 & 0.02 &    (9.3$\pm$3.82)${\times}10^{10}$ &          4.3${\times}10^{10}$ &   18.00 &  2\\
                    PG0052+251 &0.155 &       $<$ 1.74 &      &  0.55$\pm$0.12 & 0.05 &    (1.3$\pm$0.62)${\times}10^{11}$ &          3.2${\times}10^{10}$ &    2.00 &  3\\
                    PG0157+001 &0.163 &  6.71$\pm$2.44 & 0.25 &  2.44$\pm$0.16 & 0.09 &    (8.9$\pm$3.61)${\times}10^{11}$ &          5.7${\times}10^{10}$ &    8.10 &   \\
                    PG0804+761 &0.100 &       $<$ 1.75 &      &       $<$ 0.19 &      &            $<$3.8${\times}10^{10}$ &          5.6${\times}10^{10}$ &    2.00 &  2\\
                    PG0838+770 &0.131 &  1.46$\pm$0.60 & 0.17 &       $<$ 0.23 &      &    (6.1$\pm$3.09)${\times}10^{10}$ &          1.2${\times}10^{10}$ &    3.40 &  1\\
                    PG0844+349 &0.064 &  1.56$\pm$0.60 & 0.09 &  0.38$\pm$0.07 & 0.03 &    (1.0$\pm$0.44)${\times}10^{10}$ &          6.5${\times}10^{09}$ &$<$ 1.50 &  2\\
                    PG0921+525 &0.035 &       $<$ 0.59 &      &       $<$ 0.05 &      &            $<$8.6${\times}10^{08}$ &          2.2${\times}10^{09}$ &         &   \\
                    PG0923+201 &0.190 &       $<$ 0.35 &      &       $<$ 0.29 &      &            $<$9.0${\times}10^{10}$ &          5.9${\times}10^{10}$ &         &   \\
                    PG0923+129 &0.029 &  9.73$\pm$2.28 & 0.28 &  2.42$\pm$0.13 & 0.08 &    (1.3$\pm$0.51)${\times}10^{10}$ &          1.6${\times}10^{09}$ &         &   \\
                    PG0934+013 &0.050 &  2.86$\pm$0.60 & 0.26 &  0.74$\pm$0.05 & 0.08 &    (1.2$\pm$0.48)${\times}10^{10}$ &          1.8${\times}10^{09}$ &         &   \\
                    PG0946+301 &1.216 &       $<$ 0.47 &      &       $<$ 0.11 &      &            $<$1.8${\times}10^{13}$ &          1.7${\times}10^{12}$ &         &   \\
                    PG0947+396 &0.205 &       $<$ 0.38 &      &       $<$ 0.18 &      &            $<$1.2${\times}10^{11}$ &          5.0${\times}10^{10}$ &         &   \\
                    PG0953+414 &0.234 &       $<$ 1.39 &      &       $<$ 0.20 &      &            $<$3.8${\times}10^{11}$ &          7.8${\times}10^{10}$ &         &   \\
                    PG1001+054 &0.160 &       $<$ 0.38 &      &  0.17$\pm$0.03 & 0.03 &    (3.8$\pm$1.66)${\times}10^{10}$ &          2.7${\times}10^{10}$ &         &   \\
                    PG1004+130 &0.240 &       $<$ 0.58 &      &  0.20$\pm$0.05 & 0.02 &    (1.3$\pm$0.62)${\times}10^{11}$ &          6.2${\times}10^{10}$ &         &   \\
                    PG1011-040 &0.058 &       $<$ 0.56 &      &  0.50$\pm$0.04 & 0.03 &    (1.1$\pm$0.44)${\times}10^{10}$ &          3.6${\times}10^{09}$ &         &   \\
                    PG1012+008 &0.186 &       $<$ 0.61 &      &       $<$ 0.09 &      &            $<$8.2${\times}10^{10}$ &          3.6${\times}10^{10}$ &         &   \\
                    PG1022+519 &0.044 &  4.22$\pm$0.74 & 0.44 &  1.32$\pm$0.08 & 0.19 &    (1.8$\pm$0.73)${\times}10^{10}$ &          1.4${\times}10^{09}$ &         &   \\
                    PG1048+342 &0.167 &       $<$ 0.33 &      &       $<$ 0.04 &      &            $<$2.1${\times}10^{10}$ &          1.3${\times}10^{10}$ &         &   \\
                    PG1048-090 &0.344 &       $<$ 0.33 &      &       $<$ 0.06 &      &            $<$2.8${\times}10^{11}$ &          5.4${\times}10^{10}$ &         &   \\
                    PG1049-005 &0.359 &  1.17$\pm$0.38 & 0.07 &  0.17$\pm$0.07 & 0.01 &    (3.4$\pm$1.93)${\times}10^{11}$ &          2.2${\times}10^{11}$ &         &   \\
                    PG1100+772 &0.311 &       $<$ 1.04 &      &  0.29$\pm$0.08 & 0.04 &    (4.1$\pm$1.99)${\times}10^{11}$ &          1.0${\times}10^{11}$ &         &   \\
                    PG1103-006 &0.423 &       $<$ 0.18 &      &       $<$ 0.09 &      &            $<$3.6${\times}10^{11}$ &          1.3${\times}10^{11}$ &         &   \\
                    PG1114+445 &0.143 &       $<$ 0.40 &      &       $<$ 0.11 &      &            $<$5.2${\times}10^{10}$ &          4.4${\times}10^{10}$ &         &   \\
                    PG1115+407 &0.154 &  2.55$\pm$0.33 & 0.28 &  0.46$\pm$0.03 & 0.08 &    (1.1$\pm$0.46)${\times}10^{11}$ &          2.1${\times}10^{10}$ &         &   \\
                    PG1116+215 &0.176 &       $<$ 3.32 &      &       $<$ 0.25 &      &            $<$2.3${\times}10^{11}$ &          1.1${\times}10^{11}$ &         &   \\
                    PG1119+120 &0.050 &  2.26$\pm$0.89 & 0.06 &  0.80$\pm$0.09 & 0.03 &    (1.3$\pm$0.53)${\times}10^{10}$ &          5.0${\times}10^{09}$ &    4.50 &  1\\
                    PG1121+422 &0.225 &       $<$ 0.35 &      &       $<$ 0.09 &      &            $<$1.2${\times}10^{11}$ &          3.0${\times}10^{10}$ &         &   \\
                    PG1126-041 &0.060 &       $<$ 1.23 &      &  1.35$\pm$0.36 & 0.04 &    (3.6$\pm$1.75)${\times}10^{10}$ &          1.6${\times}10^{10}$ &$<$ 2.60 &  1\\
                    PG1149-110 &0.049 &       $<$ 0.64 &      &       $<$ 0.10 &      &            $<$3.5${\times}10^{09}$ &          2.2${\times}10^{09}$ &         &   \\
                    PG1151+117 &0.176 &       $<$ 3.30 &      &       $<$ 0.48 &      &            $<$5.2${\times}10^{11}$ &          1.9${\times}10^{10}$ &         &   \\
                    PG1202+281 &0.165 &  1.41$\pm$0.48 & 0.14 &  0.37$\pm$0.05 & 0.04 &    (1.0$\pm$0.44)${\times}10^{11}$ &          2.5${\times}10^{10}$ &$<$ 2.40 &  1\\
                    PG1211+143 &0.080 &       $<$ 1.82 &      &       $<$ 0.15 &      &            $<$1.9${\times}10^{10}$ &          2.8${\times}10^{10}$ &$<$ 1.50 &  2\\
                    PG1216+069 &0.331 &       $<$ 0.34 &      &       $<$ 0.05 &      &            $<$1.9${\times}10^{11}$ &          8.9${\times}10^{10}$ &         &   \\
                    PG1226+023 &0.158 &       $<$ 2.16 &      &       $<$ 0.16 &      &            $<$1.0${\times}10^{11}$ &          3.7${\times}10^{11}$ &         &   \\
                    PG1229+204 &0.063 &       $<$ 0.54 &      &  0.38$\pm$0.12 & 0.02 &    (9.8$\pm$4.94)${\times}10^{09}$ &          7.0${\times}10^{09}$ &    2.40 &  2\\
                    PG1244+026 &0.048 &  1.76$\pm$0.86 & 0.14 &  0.51$\pm$0.04 & 0.04 &    (7.0$\pm$2.86)${\times}10^{09}$ &          1.8${\times}10^{09}$ &         &   \\
                    PG1259+593 &0.477 &       $<$ 0.16 &      &       $<$ 0.04 &      &            $<$3.8${\times}10^{11}$ &          2.7${\times}10^{11}$ &         &   \\
                    PG1302-102 &0.278 &       $<$ 0.51 &      &       $<$ 0.14 &      &            $<$3.7${\times}10^{11}$ &          1.0${\times}10^{11}$ &         &   \\
                    PG1307+085 &0.155 &       $<$ 3.47 &      &       $<$ 0.43 &      &            $<$3.2${\times}10^{11}$ &          2.6${\times}10^{10}$ &         &   \\
                    PG1309+355 &0.184 &       $<$ 3.17 &      &       $<$ 0.36 &      &            $<$3.9${\times}10^{11}$ &          4.4${\times}10^{10}$ &$<$ 0.61 &  3\\
                    PG1310-108 &0.034 &  2.40$\pm$0.86 & 0.11 &  0.18$\pm$0.03 & 0.01 &    (1.0$\pm$0.44)${\times}10^{09}$ &          1.3${\times}10^{09}$ &         &   \\
                    PG1322+659 &0.168 &  0.72$\pm$0.30 & 0.07 &  0.20$\pm$0.02 & 0.03 &    (5.3$\pm$2.20)${\times}10^{10}$ &          2.9${\times}10^{10}$ &         &   \\
                    PG1341+258 &0.087 &  0.45$\pm$0.21 & 0.06 &  0.11$\pm$0.02 & 0.02 &    (5.3$\pm$2.39)${\times}10^{09}$ &          4.7${\times}10^{09}$ &         &   \\
                    PG1351+236 &0.055 &  7.54$\pm$1.05 & 0.87 &  2.75$\pm$0.12 & 0.44 &    (6.7$\pm$2.71)${\times}10^{10}$ &          1.6${\times}10^{09}$ &         &   \\
                    PG1351+640 &0.088 &  3.12$\pm$6.54 & 0.09 &  1.29$\pm$0.15 & 0.03 &    (9.3$\pm$3.89)${\times}10^{10}$ &          2.2${\times}10^{10}$ &    4.00 &  2\\
                    PG1352+183 &0.152 &       $<$14.14 &      &       $<$ 2.60 &      &            $<$2.4${\times}10^{12}$ &          1.7${\times}10^{10}$ &         &   \\
                    PG1354+213 &0.300 &       $<$ 0.27 &      &       $<$ 0.06 &      &            $<$1.8${\times}10^{11}$ &          4.2${\times}10^{10}$ &         &   \\
                    PG1402+261 &0.164 &       $<$ 1.59 &      &       $<$ 0.22 &      &            $<$1.6${\times}10^{11}$ &          6.8${\times}10^{10}$ &$<$ 2.00 &  1\\
                    PG1404+226 &0.098 &  0.88$\pm$0.37 & 0.14 &  0.25$\pm$0.02 & 0.05 &    (1.7$\pm$0.71)${\times}10^{10}$ &          5.1${\times}10^{09}$ &         &  2\\
                    PG1411+442 &0.089 &                &      &  0.31$\pm$0.04 & 0.01 &    (1.8$\pm$0.74)${\times}10^{10}$ &                               &$<$ 1.80 &  2\\
                    PG1415+451 &0.113 &  1.67$\pm$0.30 & 0.14 &  0.86$\pm$0.06 & 0.10 &    (1.1$\pm$0.43)${\times}10^{11}$ &          1.3${\times}10^{10}$ &    3.30 &  1\\
                    PG1416-129 &0.129 &       $<$ 0.56 &      &       $<$ 0.15 &      &            $<$5.8${\times}10^{10}$ &          8.5${\times}10^{09}$ &         &   \\
                    PG1425+267 &0.366 &       $<$ 0.45 &      &       $<$ 0.06 &      &            $<$3.1${\times}10^{11}$ &          1.1${\times}10^{11}$ &         &   \\
                    PG1426+015 &0.086 &  1.19$\pm$0.64 & 0.03 &  0.31$\pm$0.06 & 0.01 &    (1.7$\pm$0.73)${\times}10^{10}$ &          2.4${\times}10^{10}$ &    3.60 &  2\\
                    PG1427+480 &0.221 &       $<$ 0.28 &      &       $<$ 0.03 &      &            $<$3.4${\times}10^{10}$ &          2.5${\times}10^{10}$ &         &   \\
                    PG1435-067 &0.126 &       $<$ 0.44 &      &       $<$ 0.19 &      &            $<$4.3${\times}10^{10}$ &          1.7${\times}10^{10}$ &         &   \\
                    PG1440+356 &0.079 &  6.74$\pm$2.89 & 0.20 &  2.27$\pm$0.13 & 0.10 &    (1.3$\pm$0.53)${\times}10^{11}$ &          2.0${\times}10^{10}$ &    9.00 &  2\\
                    PG1444+407 &0.267 &  0.38$\pm$0.28 & 0.03 &       $<$ 0.15 &      &    (7.8$\pm$6.19)${\times}10^{10}$ &          8.5${\times}10^{10}$ &    0.71 &  3\\
                    PG1448+273 &0.065 &  1.98$\pm$0.59 & 0.11 &  0.94$\pm$0.06 & 0.07 &    (3.0$\pm$1.22)${\times}10^{10}$ &          5.7${\times}10^{09}$ &         &   \\
                    PG1501+106 &0.036 &       $<$ 1.70 &      &       $<$ 0.38 &      &            $<$7.9${\times}10^{09}$ &          4.2${\times}10^{09}$ &         &   \\
                    PG1512+370 &0.370 &       $<$ 0.22 &      &       $<$ 0.07 &      &            $<$3.1${\times}10^{11}$ &          8.8${\times}10^{10}$ &         &   \\
                    PG1519+226 &0.137 &  0.59$\pm$0.21 & 0.04 &  0.21$\pm$0.02 & 0.02 &    (3.3$\pm$1.37)${\times}10^{10}$ &          2.7${\times}10^{10}$ &         &   \\
                    PG1534+580 &0.029 &  1.45$\pm$0.72 & 0.05 &  0.44$\pm$0.08 & 0.02 &    (2.0$\pm$0.88)${\times}10^{09}$ &          1.5${\times}10^{09}$ &         &   \\
                    PG1535+547 &0.038 &  0.62$\pm$0.22 & 0.02 &  0.08$\pm$0.03 & 0.01 &    (5.6$\pm$2.87)${\times}10^{08}$ &          3.1${\times}10^{09}$ &         &   \\
                    PG1543+489 &0.399 &       $<$ 0.34 &      &       $<$ 0.26 &      &            $<$6.3${\times}10^{11}$ &          2.4${\times}10^{11}$ &         &   \\
                    PG1545+210 &0.264 &       $<$ 1.75 &      &       $<$ 0.17 &      &            $<$4.1${\times}10^{11}$ &          5.8${\times}10^{10}$ &$<$ 0.96 &  3\\
                    PG1552+085 &0.119 &       $<$ 0.30 &      &  0.11$\pm$0.02 & 0.02 &    (1.1$\pm$0.46)${\times}10^{10}$ &          1.0${\times}10^{10}$ &         &   \\
                    PG1612+261 &0.130 &       $<$ 0.46 &      &  0.38$\pm$0.22 & 0.03 &    (5.8$\pm$4.17)${\times}10^{10}$ &          2.1${\times}10^{10}$ &         &   \\
                    PG1613+658 &0.129 &  3.02$\pm$1.87 & 0.08 &  0.77$\pm$0.09 & 0.03 &    (1.3$\pm$0.53)${\times}10^{11}$ &          5.5${\times}10^{10}$ &    8.50 &  1\\
                    PG1617+175 &0.112 &       $<$ 0.45 &      &       $<$ 0.48 &      &            $<$3.2${\times}10^{10}$ &          1.9${\times}10^{10}$ &         &   \\
                    PG1626+554 &0.133 &       $<$ 0.47 &      &       $<$ 0.09 &      &            $<$3.2${\times}10^{10}$ &          1.3${\times}10^{10}$ &         &   \\
                    PG1634+706 &1.334 &       $<$ 0.52 &      &       $<$ 0.11 &      &            $<$2.4${\times}10^{13}$ &          9.4${\times}10^{09}$ &         &   \\
                    PG1700+518 &0.292 &       $<$ 5.70 &      &       $<$ 0.20 &      &            $<$6.5${\times}10^{11}$ &          3.2${\times}10^{11}$ &         &   \\
                    PG1704+608 &0.371 &       $<$ 1.04 &      &       $<$ 0.11 &      &            $<$6.5${\times}10^{11}$ &          2.6${\times}10^{11}$ &         &   \\
                    PG2112+059 &0.466 &       $<$ 0.24 &      &  0.27$\pm$0.05 & 0.02 &    (1.2$\pm$0.52)${\times}10^{12}$ &          4.8${\times}10^{11}$ &         &   \\
                    PG2130+099 &0.062 &  4.20$\pm$1.29 & 0.06 &  0.55$\pm$0.21 & 0.01 &    (1.5$\pm$0.83)${\times}10^{10}$ &          2.1${\times}10^{10}$ &    4.30 &  2\\
                    PG2209+184 &0.070 &  1.32$\pm$0.37 & 0.20 &  0.29$\pm$0.03 & 0.06 &    (9.1$\pm$3.74)${\times}10^{09}$ &          3.0${\times}10^{09}$ &         &   \\
                    PG2214+139 &0.065 &       $<$ 0.81 &      &       $<$ 0.28 &      &            $<$1.7${\times}10^{10}$ &          1.6${\times}10^{10}$ &    1.60 &  2\\
                    PG2233+134 &0.325 &       $<$ 1.44 &      &       $<$ 0.15 &      &            $<$6.6${\times}10^{11}$ &          9.4${\times}10^{10}$ &         &   \\
                    PG2251+113 &0.325 &       $<$ 0.55 &      &       $<$ 0.26 &      &            $<$6.3${\times}10^{11}$ &          1.5${\times}10^{11}$ &         &   \\
                    PG2304+042 &0.042 &       $<$ 0.46 &      &       $<$ 0.05 &      &            $<$5.6${\times}10^{09}$ &          8.8${\times}10^{08}$ &         &   \\
                    PG2308+098 &0.433 &       $<$ 0.27 &      &       $<$ 0.06 &      &            $<$5.4${\times}10^{11}$ &          1.4${\times}10^{11}$ &         &   \\
                    PG2349-014 &0.174 &       $<$ 0.47 &      &  0.41$\pm$0.10 & 0.05 &    (1.3$\pm$0.61)${\times}10^{11}$ &          4.6${\times}10^{10}$ &    3.20 &  3\\
      2MASSJ000703.61+155423.8 &0.114 &  3.00$\pm$0.71 & 0.32 &  1.00$\pm$0.10 & 0.14 &    (1.3$\pm$0.52)${\times}10^{11}$ &          9.7${\times}10^{09}$ &         &   \\
      2MASSJ005055.70+293328.1 &0.136 &  1.76$\pm$0.34 & 0.19 &  0.33$\pm$0.09 & 0.05 &    (5.6$\pm$2.71)${\times}10^{10}$ &          1.5${\times}10^{10}$ &         &   \\
      2MASSJ010835.16+214818.6 &0.285 &       $<$ 1.25 &      &       $<$ 0.21 &      &            $<$6.6${\times}10^{11}$ &          1.1${\times}10^{11}$ &         &   \\
      2MASSJ015721.05+171248.4 &0.213 &  2.02$\pm$0.46 & 0.33 &  0.58$\pm$0.12 & 0.16 &    (3.5$\pm$1.56)${\times}10^{11}$ &          3.0${\times}10^{10}$ &         &   \\
      2MASSJ022150.60+132741.0 &0.140 &       $<$ 3.31 &      &       $<$ 0.39 &      &            $<$2.1${\times}10^{11}$ &          2.5${\times}10^{10}$ &         &   \\
      2MASSJ023430.64+243835.5 &0.310 &       $<$ 1.16 &      &       $<$ 0.34 &      &            $<$1.3${\times}10^{12}$ &          6.4${\times}10^{10}$ &         &   \\
      2MASSJ034857.64+125547.3 &0.210 &       $<$ 1.56 &      &       $<$ 0.33 &      &            $<$5.2${\times}10^{11}$ &          2.3${\times}10^{11}$ &         &   \\
      2MASSJ091848.63+211717.1 &0.149 &       $<$ 1.27 &      &  0.45$\pm$0.25 & 0.04 &    (1.0$\pm$0.69)${\times}10^{11}$ &          3.2${\times}10^{10}$ &         &   \\
      2MASSJ095504.56+170556.1 &0.139 &       $<$ 1.10 &      &       $<$ 0.28 &      &            $<$1.4${\times}10^{11}$ &          9.5${\times}10^{09}$ &         &   \\
      2MASSJ102724.95+121920.4 &0.231 &       $<$ 1.49 &      &       $<$ 0.36 &      &            $<$7.1${\times}10^{11}$ &          7.0${\times}10^{10}$ &         &   \\
      2MASSJ105144.25+353930.7 &0.158 &       $<$ 0.99 &      &       $<$ 0.24 &      &            $<$1.7${\times}10^{11}$ &          1.0${\times}10^{10}$ &         &   \\
      2MASSJ125807.46+232921.5 &0.259 &  1.54$\pm$0.78 & 0.09 &       $<$ 0.16 &      &    (3.9$\pm$2.29)${\times}10^{11}$ &          9.4${\times}10^{10}$ &         &   \\
      2MASSJ130005.35+163214.8 &0.080 &       $<$ 4.82 &      &       $<$ 1.05 &      &            $<$1.7${\times}10^{11}$ &          2.2${\times}10^{10}$ &         &   \\
      2MASSJ130700.66+233805.0 &0.275 &  9.27$\pm$1.58 & 0.57 &       $<$ 0.27 &      &    (3.9$\pm$1.33)${\times}10^{12}$ &          2.2${\times}10^{11}$ &         &   \\
      2MASSJ140251.22+263117.5 &0.187 &       $<$ 1.46 &      &       $<$ 0.58 &      &            $<$4.4${\times}10^{11}$ &          2.6${\times}10^{10}$ &         &   \\
      2MASSJ145331.51+135358.7 &0.139 & 10.31$\pm$2.81 & 0.51 &  1.10$\pm$0.70 & 0.16 &    (2.3$\pm$1.76)${\times}10^{11}$ &          4.0${\times}10^{10}$ &         &   \\
      2MASSJ150113.21+232908.3 &0.258 &       $<$ 1.31 &      &  0.19$\pm$0.08 & 0.03 &    (1.4$\pm$0.82)${\times}10^{11}$ &          5.0${\times}10^{10}$ &         &   \\
      2MASSJ151653.24+190048.4 &0.190 &       $<$ 2.22 &      &       $<$ 0.57 &      &            $<$7.2${\times}10^{11}$ &          1.6${\times}10^{11}$ &         &   \\
      2MASSJ163700.22+222114.0 &0.211 &  2.81$\pm$0.61 & 0.60 &  0.51$\pm$0.04 & 0.15 &    (2.7$\pm$1.10)${\times}10^{11}$ &          2.0${\times}10^{10}$ &         &   \\
      2MASSJ165939.77+183436.9 &0.170 &  3.43$\pm$1.21 & 0.17 &  0.65$\pm$0.16 & 0.04 &    (2.2$\pm$1.01)${\times}10^{11}$ &          4.7${\times}10^{10}$ &         &   \\
      2MASSJ171442.77+260248.5 &0.163 &  1.19$\pm$0.34 & 0.20 &  0.32$\pm$0.07 & 0.08 &    (8.0$\pm$3.60)${\times}10^{10}$ &          1.7${\times}10^{10}$ &         &   \\
      2MASSJ222202.22+195231.5 &0.366 &       $<$ 0.97 &      &       $<$ 0.09 &      &            $<$5.0${\times}10^{11}$ &          2.7${\times}10^{11}$ &         &   \\
      2MASSJ222221.12+195947.4 &0.211 &       $<$ 1.05 &      &       $<$ 0.14 &      &            $<$1.8${\times}10^{11}$ &          4.3${\times}10^{10}$ &         &   \\
      2MASSJ222554.27+195837.0 &0.147 &  1.97$\pm$0.33 & 0.22 &       $<$ 0.21 &      &    (1.2$\pm$0.40)${\times}10^{11}$ &          1.6${\times}10^{10}$ &         &   \\
      2MASSJ234449.57+122143.4 &0.199 &       $<$ 1.25 &      &       $<$ 0.15 &      &            $<$1.8${\times}10^{11}$ &          3.5${\times}10^{10}$ &         &   \\
                         3C6.1 &0.840 &                &      &       $<$ 0.05 &      &            $<$2.5${\times}10^{12}$ &                               &         &   \\
                          3C15 &0.073 &       $<$ 0.52 &      &       $<$ 0.03 &      &            $<$2.4${\times}10^{09}$ &          1.0${\times}10^{09}$ &         &   \\
                          3C20 &0.174 &       $<$ 1.20 &      &       $<$ 0.24 &      &            $<$2.1${\times}10^{11}$ &          4.2${\times}10^{09}$ &         &   \\
                          3C22 &0.936 &                &      &       $<$ 0.03 &      &            $<$1.5${\times}10^{12}$ &                               &         &   \\
                          3C28 &0.195 &       $<$ 0.30 &      &       $<$ 0.06 &      &            $<$5.3${\times}10^{10}$ &          8.6${\times}10^{08}$ &         &   \\
                          3C29 &0.045 &       $<$ 0.75 &      &       $<$ 0.02 &      &            $<$6.6${\times}10^{08}$ &          2.3${\times}10^{08}$ &         &   \\
                          3C33 &0.059 &       $<$ 0.60 &      &       $<$ 0.15 &      &            $<$9.1${\times}10^{09}$ &          2.7${\times}10^{09}$ &         &   \\
                        3C33.1 &0.180 &       $<$ 1.62 &      &       $<$ 0.39 &      &            $<$4.0${\times}10^{11}$ &          1.2${\times}10^{10}$ &         &   \\
                          3C47 &0.425 &       $<$ 0.28 &      &       $<$ 0.06 &      &            $<$4.1${\times}10^{11}$ &          1.1${\times}10^{11}$ &         &   \\
                          3C48 &0.367 &       $<$ 4.55 &      &       $<$ 0.53 &      &            $<$4.0${\times}10^{12}$ &          2.5${\times}10^{11}$ &    2.00 &  4\\
                          3C55 &0.734 &       $<$ 0.18 &      &       $<$ 0.07 &      &            $<$1.7${\times}10^{12}$ &          8.9${\times}10^{10}$ &         &   \\
                        3C61.1 &0.187 &       $<$ 0.23 &      &       $<$ 0.05 &      &            $<$4.2${\times}10^{10}$ &          8.3${\times}10^{08}$ &         &   \\
                          3C65 &1.176 &       $<$ 0.42 &      &       $<$ 0.08 &      &            $<$1.2${\times}10^{13}$ &          8.1${\times}10^{09}$ &         &   \\
                          3C75 &0.023 &       $<$ 0.33 &      &       $<$ 0.01 &      &            $<$7.6${\times}10^{07}$ &          2.6${\times}10^{07}$ &         &   \\
                        3C76.1 &0.032 &       $<$ 0.43 &      &       $<$ 0.05 &      &            $<$6.3${\times}10^{08}$ &          9.4${\times}10^{07}$ &         &   \\
                          3C79 &0.255 &       $<$ 1.02 &      &       $<$ 0.17 &      &            $<$3.9${\times}10^{11}$ &          3.0${\times}10^{10}$ &         &   \\
                        3C83.1 &0.025 &       $<$ 0.36 &      &  0.16$\pm$0.02 & 0.09 &    (4.4$\pm$1.84)${\times}10^{08}$ &          4.6${\times}10^{08}$ &         &   \\
                          3C84 &0.017 &                &      &  4.11$\pm$1.19 & 0.02 &    (7.1$\pm$3.53)${\times}10^{09}$ &          1.6${\times}10^{09}$ &         &   \\
                         3C109 &0.305 &       $<$ 1.75 &      &       $<$ 0.26 &      &            $<$1.1${\times}10^{12}$ &          2.2${\times}10^{11}$ &         &   \\
                         3C123 &0.217 &       $<$ 0.61 &      &       $<$ 0.05 &      &            $<$5.6${\times}10^{10}$ &          1.3${\times}10^{09}$ &         &   \\
                         3C129 &0.020 &       $<$ 0.36 &      &  0.07$\pm$0.01 & 0.06 &    (1.3$\pm$0.56)${\times}10^{08}$ &          1.2${\times}10^{08}$ &         &   \\
                         3C138 &0.759 &                &      &       $<$ 0.03 &      &            $<$1.0${\times}10^{12}$ &                               &         &   \\
                         3C147 &0.545 &                &      &       $<$ 0.05 &      &            $<$6.6${\times}10^{11}$ &                               &         &   \\
                         3C153 &0.276 &       $<$ 0.40 &      &       $<$ 0.04 &      &            $<$8.2${\times}10^{10}$ &          4.7${\times}10^{08}$ &         &   \\
                         3C172 &0.519 &       $<$ 0.17 &      &       $<$ 0.05 &      &            $<$5.8${\times}10^{11}$ &          5.5${\times}10^{09}$ &         &   \\
                       3C173.1 &0.292 &       $<$ 0.31 &      &       $<$ 0.03 &      &            $<$6.6${\times}10^{10}$ &          1.6${\times}10^{09}$ &         &   \\
                         3C175 &0.770 &       $<$ 0.17 &      &       $<$ 0.03 &      &            $<$1.0${\times}10^{12}$ &          2.8${\times}10^{11}$ &         &   \\
                         3C184 &0.994 &                &      &       $<$ 0.06 &      &            $<$5.7${\times}10^{12}$ &                               &         &   \\
                         3C192 &0.059 &       $<$ 0.43 &      &       $<$ 0.05 &      &            $<$2.8${\times}10^{09}$ &          2.1${\times}10^{08}$ &         &   \\
                         3C196 &0.871 &       $<$ 0.14 &      &       $<$ 0.04 &      &            $<$2.1${\times}10^{12}$ &          3.3${\times}10^{11}$ &         &   \\
                         3C200 &0.458 &                &      &       $<$ 0.09 &      &            $<$9.4${\times}10^{11}$ &                               &         &   \\
                         3C216 &0.670 &       $<$ 0.28 &      &       $<$ 0.12 &      &            $<$2.3${\times}10^{12}$ &          2.5${\times}10^{11}$ &         &   \\
                         3C219 &0.174 &       $<$ 0.30 &      &       $<$ 0.10 &      &            $<$5.9${\times}10^{10}$ &          6.0${\times}10^{09}$ &         &   \\
                       3C220.1 &0.610 &       $<$ 0.29 &      &       $<$ 0.10 &      &            $<$1.8${\times}10^{12}$ &          1.7${\times}10^{10}$ &         &   \\
                       3C220.3 &0.680 &       $<$ 1.05 &      &       $<$ 0.03 &      &            $<$6.9${\times}10^{11}$ &          1.6${\times}10^{10}$ &         &   \\
                         3C234 &0.184 &       $<$ 0.90 &      &       $<$ 0.24 &      &            $<$2.4${\times}10^{11}$ &          1.2${\times}10^{11}$ &         &   \\
                       3C244.1 &0.428 &       $<$ 0.21 &      &       $<$ 0.04 &      &            $<$3.3${\times}10^{11}$ &          2.5${\times}10^{10}$ &         &   \\
                       3C249.1 &0.311 &       $<$ 1.04 &      &  0.29$\pm$0.10 & 0.04 &    (4.1$\pm$2.16)${\times}10^{11}$ &          1.0${\times}10^{11}$ &         &   \\
                         3C263 &0.646 &       $<$ 0.14 &      &       $<$ 0.07 &      &            $<$9.4${\times}10^{11}$ &          3.6${\times}10^{11}$ &         &   \\
                       3C263.1 &0.824 &       $<$ 0.10 &      &       $<$ 0.16 &      &            $<$1.2${\times}10^{12}$ &          1.6${\times}10^{10}$ &         &   \\
                         3C265 &0.811 &  0.62$\pm$0.23 & 0.24 &       $<$ 0.30 &      &    (3.4$\pm$1.62)${\times}10^{12}$ &          2.6${\times}10^{11}$ &         &   \\
                       3C268.1 &0.970 &       $<$ 0.15 &      &       $<$ 0.08 &      &            $<$3.2${\times}10^{12}$ &          2.2${\times}10^{10}$ &         &   \\
                         3C270 &0.007 &                &      &  0.60$\pm$0.04 & 0.09 &    (1.4$\pm$0.56)${\times}10^{08}$ &          5.4${\times}10^{07}$ &         &   \\
                         3C272 &0.944 &                &      &       $<$ 0.02 &      &            $<$1.1${\times}10^{12}$ &                               &         &   \\
                       3C272.1 &0.003 &                &      &  1.70$\pm$0.12 & 0.33 &    (2.0$\pm$0.04)${\times}10^{09}$ &          3.0${\times}10^{07}$ &         &   \\
                         3C273 &0.158 &       $<$ 2.16 &      &       $<$ 0.16 &      &            $<$1.0${\times}10^{11}$ &          3.7${\times}10^{11}$ &         &   \\
                         3C274 &0.004 &                &      &       $<$ 0.97 &      &            $<$2.3${\times}10^{08}$ &          4.3${\times}10^{07}$ &$<$ 11.7 &  5\\
                       3C274.1 &0.422 &       $<$ 0.19 &      &       $<$ 0.08 &      &            $<$3.7${\times}10^{11}$ &          3.0${\times}10^{09}$ &         &   \\
                       3C275.1 &0.555 &       $<$ 0.15 &      &  0.09$\pm$0.02 & 0.09 &    (5.1$\pm$2.29)${\times}10^{11}$ &          5.2${\times}10^{10}$ &         &   \\
                         3C280 &0.996 &       $<$ 0.09 &      &       $<$ 0.09 &      &            $<$1.9${\times}10^{12}$ &          2.3${\times}10^{11}$ &         &   \\
                         3C292 &0.710 &                &      &       $<$ 0.09 &      &            $<$3.4${\times}10^{12}$ &                               &         &   \\
                         3C293 &0.045 &  3.96$\pm$0.70 & 0.62 &  1.27$\pm$0.10 & 0.41 &    (1.7$\pm$0.71)${\times}10^{10}$ &          9.2${\times}10^{08}$ &         &   \\
                         3C295 &0.464 &       $<$ 0.13 &      &       $<$ 0.24 &      &            $<$3.1${\times}10^{11}$ &          3.5${\times}10^{09}$ &         &   \\
                         3C298 &1.436 &       $<$ 0.30 &      &       $<$ 0.07 &      &            $<$1.7${\times}10^{13}$ &          1.2${\times}10^{12}$ &         &   \\
                         3C300 &0.270 &       $<$ 0.34 &      &       $<$ 0.06 &      &            $<$1.2${\times}10^{11}$ &          1.5${\times}10^{09}$ &         &   \\
                       3C303.1 &0.267 &       $<$ 0.38 &      &  0.09$\pm$0.02 & 0.18 &    (6.9$\pm$3.21)${\times}10^{10}$ &          5.3${\times}10^{09}$ &         &   \\
                       3C309.1 &0.905 &       $<$ 0.11 &      &       $<$ 0.03 &      &            $<$1.9${\times}10^{12}$ &          3.3${\times}10^{11}$ &         &   \\
                         3C310 &0.053 &       $<$ 0.30 &      &       $<$ 0.03 &      &            $<$1.3${\times}10^{09}$ &          1.5${\times}10^{08}$ &         &   \\
                         3C315 &0.108 &       $<$ 0.44 &      &  0.17$\pm$0.02 & 0.50 &    (1.4$\pm$0.61)${\times}10^{10}$ &          5.4${\times}10^{08}$ &         &   \\
                         3C318 &1.574 &       $<$ 0.51 &      &       $<$ 0.07 &      &            $<$2.3${\times}10^{13}$ &          2.8${\times}10^{11}$ &         &   \\
                         3C319 &0.192 &       $<$ 0.23 &      &       $<$ 0.08 &      &            $<$5.6${\times}10^{10}$ &                               &         &   \\
                         3C321 &0.096 &  6.51$\pm$1.04 & 0.49 &       $<$ 0.28 &      &    (1.7$\pm$0.57)${\times}10^{11}$ &          6.1${\times}10^{09}$ &$<$ 4.70 &  5\\
                       3C323.1 &0.264 &       $<$ 1.75 &      &       $<$ 0.17 &      &            $<$4.1${\times}10^{11}$ &          5.8${\times}10^{10}$ &         &   \\
                         3C325 &1.135 &       $<$ 0.10 &      &       $<$ 0.04 &      &            $<$3.3${\times}10^{12}$ &          9.3${\times}10^{10}$ &         &   \\
                         3C326 &0.089 &       $<$ 0.62 &      &       $<$ 0.11 &      &            $<$1.7${\times}10^{10}$ &          3.2${\times}10^{08}$ &         &   \\
                         3C330 &0.550 &  0.25$\pm$0.07 & 0.29 &       $<$ 0.02 &      &    (3.8$\pm$1.54)${\times}10^{11}$ &          2.8${\times}10^{10}$ &         &   \\
                         3C334 &0.555 &  0.58$\pm$0.21 & 0.17 &       $<$ 0.03 &      &    (1.1$\pm$0.50)${\times}10^{12}$ &          1.5${\times}10^{11}$ &         &   \\
                         3C336 &0.927 &                &      &       $<$ 0.08 &      &            $<$6.6${\times}10^{12}$ &                               &         &   \\
                         3C337 &0.635 &                &      &       $<$ 0.05 &      &            $<$1.2${\times}10^{12}$ &                               &         &   \\
                         3C340 &0.775 &                &      &       $<$ 0.03 &      &            $<$1.1${\times}10^{12}$ &                               &         &   \\
                         3C343 &0.988 &                &      &       $<$ 0.04 &      &            $<$3.2${\times}10^{12}$ &                               &         &   \\
                       3C343.1 &0.750 &                &      &       $<$ 0.02 &      &            $<$7.2${\times}10^{11}$ &                               &         &   \\
                         3C348 &0.154 &       $<$ 0.81 &      &       $<$ 0.19 &      &            $<$1.1${\times}10^{11}$ &          8.3${\times}10^{08}$ &         &   \\
                         3C351 &0.371 &       $<$ 1.04 &      &       $<$ 0.11 &      &            $<$6.5${\times}10^{11}$ &          2.6${\times}10^{11}$ &         &   \\
                         3C352 &0.806 &                &      &       $<$ 0.05 &      &            $<$2.3${\times}10^{12}$ &                               &         &   \\
                         3C356 &1.079 &       $<$ 0.17 &      &       $<$ 0.06 &      &            $<$5.2${\times}10^{12}$ &          8.0${\times}10^{10}$ &         &   \\
                         3C371 &0.051 &       $<$ 2.12 &      &       $<$ 0.13 &      &            $<$5.1${\times}10^{09}$ &          8.6${\times}10^{09}$ &         &   \\
                         3C380 &0.692 &       $<$ 0.17 &      &       $<$ 0.09 &      &            $<$1.3${\times}10^{12}$ &          3.5${\times}10^{11}$ &         &   \\
                         3C381 &0.160 &       $<$ 0.48 &      &       $<$ 0.05 &      &            $<$3.0${\times}10^{10}$ &          1.5${\times}10^{10}$ &         &   \\
                         3C382 &0.057 &       $<$ 0.97 &      &       $<$ 0.12 &      &            $<$6.3${\times}10^{09}$ &          1.7${\times}10^{10}$ &         &   \\
                         3C386 &0.016 &                &      &       $<$ 0.04 &      &            $<$1.4${\times}10^{08}$ &          9.0${\times}10^{07}$ &         &   \\
                         3C388 &0.091 &       $<$ 0.36 &      &       $<$ 0.06 &      &            $<$8.8${\times}10^{09}$ &          5.9${\times}10^{08}$ &         &   \\
                       3C390.3 &0.056 &       $<$ 0.63 &      &       $<$ 0.16 &      &            $<$7.9${\times}10^{09}$ &          8.3${\times}10^{09}$ &$<$ 10.3 &  5\\
                         3C401 &0.201 &       $<$ 0.27 &      &       $<$ 0.03 &      &            $<$2.9${\times}10^{10}$ &          1.1${\times}10^{09}$ &         &   \\
                       3C403.1 &0.055 &       $<$ 0.25 &      &       $<$ 0.02 &      &            $<$7.1${\times}10^{08}$ &          1.7${\times}10^{08}$ &         &   \\
                         3C405 &0.056 &       $<$ 3.28 &      &       $<$ 0.55 &      &            $<$3.3${\times}10^{10}$ &          3.5${\times}10^{09}$ &$<$ 1.90 &  5\\
                       3C427.1 &0.572 &       $<$ 0.22 &      &       $<$ 0.03 &      &            $<$5.6${\times}10^{11}$ &          3.4${\times}10^{09}$ &         &   \\
                         3C433 &0.101 &       $<$ 0.83 &      &       $<$ 0.22 &      &            $<$4.9${\times}10^{10}$ &          2.0${\times}10^{10}$ &         &   \\
                         3C436 &0.214 &       $<$ 0.40 &      &       $<$ 0.05 &      &            $<$5.4${\times}10^{10}$ &          1.1${\times}10^{09}$ &         &   \\
                         3C438 &0.290 &       $<$ 0.35 &      &       $<$ 0.04 &      &            $<$9.3${\times}10^{10}$ &          1.5${\times}10^{09}$ &         &   \\
                         3C441 &0.708 &                &      &       $<$ 0.06 &      &            $<$1.9${\times}10^{12}$ &                               &         &   \\
                         3C445 &0.056 &       $<$ 1.48 &      &       $<$ 0.30 &      &            $<$1.7${\times}10^{10}$ &          1.7${\times}10^{10}$ &         &   \\
                         3C452 &0.081 &       $<$ 0.50 &      &       $<$ 0.07 &      &            $<$7.9${\times}10^{09}$ &          2.7${\times}10^{09}$ &         &   \\
                         3C465 &0.030 &       $<$ 0.86 &      &       $<$ 0.21 &      &            $<$2.6${\times}10^{09}$ &          3.5${\times}10^{08}$ &         &   \\
\enddata
\tablecomments{ 
Column   (1):  Sources.   Column  (2):   Redshift.  Column   (3):  The
observed-frame 7.7 $\mu$m aromatic flux  in the unit of 10$^{-13}$ erg
s$^{-1}$ cm$^{-2}$.  Column  (4): The rest-frame EW of  7.7 $\mu$m PAH
in  the unit  of $\mu$m.  Column (5):  The observed-frame  11.3 $\mu$m
aromatic  flux  in the  unit  of  10$^{-13}$  erg s$^{-1}$  cm$^{-2}$.
Column  (6): The  rest-frame EW  of  11.3 $\mu$m  PAH in  the unit  of
$\mu$m.  Column  (7): The  star-forming IR luminosity  in the  unit of
L$_{\odot}$.   Column  (8):  The  mid-IR  luminosity in  the  unit  of
L$_{\odot}$ integrated  from 5 to  6 $\mu$m. A  factor of 22.6  can be
applied to convert it to the total IR luminosity (3-1000 $\mu$m) based
on the quasar template of \citet{Elvis94}.  Column (9): The CO flux in
the unit  of Jy km s$^{-1}$.   Column (10): Reference  for column (9).
\\      REFERENCES:     (1)\citet{Evans01};     (2)\citet{Scoville03};
(3)\citet{Casoli01}; (4)\citet{Scoville93}; (5)\citet{Evans05} }
\end{deluxetable}

\begin{deluxetable}{ccccccccccccccccc}
\tabletypesize{\scriptsize}
\tablecolumns{7}
\tablecaption{\label{SF_MIPS} The Star Formation Fraction at Three MIPS Bands as a Function of
the mid-IR Luminosity}
\tablewidth{0pt}
\tablehead{ \colhead{MIPS band} & \colhead{$\alpha$} & \colhead{$\beta$} & \colhead{Correlation}   }
\startdata

All(MIPS 24 $\mu$m)    & 0.6$\pm$ 1.3  & -0.18$\pm$0.13  & -0.22$\pm$   0.15 \\ 
All(MIPS 70 $\mu$m)    & 1.2$\pm$ 1.0  & -0.17$\pm$0.10  & -0.32$\pm$   0.17 \\
All(MIPS 160 $\mu$m)   & 2.3$\pm$ 5.5  & -0.27$\pm$0.51  & -0.15$\pm$   0.29 \\
\hline
PG(MIPS 24 $\mu$m)     & 0.2$\pm$ 1.3  & -0.15$\pm$0.13  & -0.22$\pm$   0.19 \\
PG(MIPS 70 $\mu$m)     & 2.4$\pm$ 1.6  & -0.29$\pm$0.15  & -0.43$\pm$   0.19 \\
\hline
2MASS(MIPS 24 $\mu$m)  & 0.1$\pm$12.3  & -0.11$\pm$1.15  &  0.01$\pm$   0.43 \\
2MASS(MIPS 70 $\mu$m)  & 1.2$\pm$ 6.9  & -0.14$\pm$0.65  & -0.05$\pm$   0.48 \\
\enddata
\tablecomments{ 
$Log$(Frac$_{SF}^{\rm MIPS}$) = $\alpha$ + $\beta$${\times}Log$($L_{MIR}$) 
}
\end{deluxetable}

\begin{deluxetable}{ccccccccccccccccc}
\tabletypesize{\scriptsize}
\tablecolumns{7}
\tablecaption{\label{SF_MIPS_Edd} The Star Formation Fraction at Three MIPS Bands as a Function of
the Eddington ratio}
\tablewidth{0pt}
\tablehead{ \colhead{MIPS band} & \colhead{$\alpha$} & \colhead{$\beta$} & \colhead{Correlation}   }
\startdata
PG(MIPS 24 $\mu$m)     & -1.6$\pm$ 0.3  & -0.10$\pm$0.12  & -0.18$\pm$   0.21 \\
PG(MIPS 70 $\mu$m)     & -1.3$\pm$ 0.3  & -0.32$\pm$0.12  & -0.60$\pm$   0.17 \\
\enddata
\tablecomments{ 
$Log$(Frac$_{SF}^{\rm MIPS}$) = $\alpha$ + $\beta$${\times}Log$($L_{MIR}/L_{Edd}$) 
}
\end{deluxetable}

\begin{deluxetable}{ccccccccccccccccc}
\tabletypesize{\scriptsize}
\tablecolumns{7}
\tablecaption{\label{FB_PG} Fractional Bivariate Luminosity Function for PG quasars}
\tablewidth{0pt}
\tablehead{ \colhead{ } & \multicolumn{4}{c}{ $M_{B}$(mag) }  \\
         \cline{2-5}      \\
\colhead{Log($L_{totIR}^{PAH}[L_{\odot}]$)}   & \colhead{-25.83}  & 
\colhead{-24.57}  & \colhead{-23.31}  & \colhead{-22.05}}
\startdata
   10.06 & 0/0$\pm$ 1.00 & 0/0$\pm$ 1.00 & 3/6$\pm$ 0.35 & 8/12$\pm$ 0.30 \\
   10.75 & 0/0$\pm$ 1.00 & 2/2$\pm$ 1.00 & 7/19$\pm$ 0.16 & 5/14$\pm$ 0.19 \\
   11.43 & 1/3$\pm$ 0.38 & 2/8$\pm$ 0.20 & 3/26$\pm$ 0.07 & 1/15$\pm$ 0.07 \\
   12.11 & 1/14$\pm$ 0.07 & 1/17$\pm$ 0.06 & 0/26$\pm$ 0.00 & 0/15$\pm$ 0.00 \\
\enddata
\end{deluxetable}

\begin{deluxetable}{ccccccccccccccccc}
\tabletypesize{\scriptsize}
\tablecolumns{7}
\tablecaption{\label{FB_2M} Fractional Bivariate Luminosity Function for 2MASS quasars}
\tablewidth{0pt}
\tablehead{ \colhead{ } & \multicolumn{3}{c}{ Log($L_{K}$ [L$_{\odot}$])  } \\
         \cline{2-4}      \\
\colhead{Log($L_{totIR}^{PAH}[L_{\odot}]$)}   & \colhead{ 10.88}  & 
\colhead{ 11.26}  & \colhead{ 11.64} }
\startdata
   11.06 & 4/5$\pm$ 0.54 & 3/2$\pm$ 1.37 & 0/0$\pm$ 1.00 \\
   11.70 & 1/9$\pm$ 0.12 & 2/8$\pm$ 0.20 & 0/0$\pm$ 1.00 \\
   12.33 & 0/10$\pm$ 0.00 & 1/12$\pm$ 0.09 & 0/2$\pm$ 0.00 \\
\enddata
\end{deluxetable}

\begin{deluxetable}{ccccccccccccccccc}
\tabletypesize{\scriptsize}
\tablecolumns{7}
\tablecaption{\label{FB_3C} Fractional Bivariate Luminosity Function for 3CR radio galaxies and quasars}
\tablewidth{0pt}
\tablehead{ \colhead{ } & \multicolumn{3}{c}{ Log($L_{\rm 151MHz}$[W Hz$^{-1}$sr$^{-1}$]) }  \\
         \cline{2-4}      \\
\colhead{Log($L_{totIR}^{PAH}[L_{\odot}]$)}   & \colhead{ 25.13}  & 
\colhead{ 26.23}  & \colhead{ 27.32}   }
\startdata
   10.54 & 4/9$\pm$ 0.27 & 1/3$\pm$ 0.38 & 0/1$\pm$ 0.00 \\
   12.01 & 0/10$\pm$ 0.00 & 1/21$\pm$ 0.05 & 0/9$\pm$ 0.00 \\
\enddata
\end{deluxetable}

\begin{deluxetable}{ccccccccccccccccc}
\tabletypesize{\scriptsize}
\tablecolumns{7}
\tablecaption{\label{Best_Fit} Best-fitting parameters to star-forming IR LF of PG quasars}
\tablewidth{0pt}
\tablehead{ \colhead{Object}                        & \colhead{Log($\phi^{\star}$[Mpc$^{-3}$ mag$^{-1}$]) } & 
            \colhead{Log($L^{\star}$[L$_{\odot}$])} & \colhead{$\alpha$}                                       }
\startdata
PG($M_{B}<$-21) &  -7.88$\pm$ 0.29 &     11.45$\pm$ 0.17 &   -1.18$\pm$ 0.24 \\
PG($M_{B}<$-23) & -8.37$\pm$ 0.35  &     11.49$\pm$ 0.42 &   -0.28$\pm$ 1.49 \\
\enddata
\tablecomments{The formula of luminosity function is a Schechter function: $\Phi(L)dL=
\Phi^{*}(\frac{L}{L^{*}})^{\alpha}exp(-\frac{L}{L^{*}})
\frac{dL}{{L^{*}}}$. }
\end{deluxetable}

\end{document}